\documentclass[english,english,draftcls,onecolumn]{IEEEtran}
\usepackage[T1]{fontenc}
\usepackage[latin9]{inputenc}
\usepackage{array}
\usepackage{float}
\usepackage{amsmath}
\usepackage{amsthm}
\usepackage{amssymb}
\usepackage{graphicx}
\PassOptionsToPackage{normalem}{ulem}
\usepackage{ulem}

\makeatletter

\providecommand{\tabularnewline}{\\}

  \theoremstyle{definition}
  \newtheorem{defn}{\protect\definitionname}
  \theoremstyle{plain}
  \newtheorem{lem}{\protect\lemmaname}
  \theoremstyle{plain}
  \newtheorem{thm}{\protect\theoremname}
  \theoremstyle{remark}
  \newtheorem{rem}{\protect\remarkname}
  \theoremstyle{plain}
  \newtheorem{conjecture}{\protect\conjecturename}

\usepackage[colorlinks,
            linkcolor=red,
            anchorcolor=blue,
            citecolor=green
            ]{hyperref}

\makeatother

\usepackage{babel}
\providecommand{\conjecturename}{Conjecture}
\providecommand{\definitionname}{Definition}
\providecommand{\lemmaname}{Lemma}
\providecommand{\remarkname}{Remark}
\providecommand{\theoremname}{Theorem}

\begin{document}

\title{Degrees of Freedom of the Two-User MIMO Broadcast Channel with Private and Common Messages Under Hybrid CSIT Models}

\author{Yao Wang and Mahesh K. Varanasi
\thanks{Yao Wang is with the Department of Electrical, Computer and Energy
Engineering, University of Colorado Boulder, Boulder, US, e-mail:
\protect\href{mailto:yao.wang-2@colorado.edu}{yao.wang-2@colorado.edu}.%
}%
\thanks{Mahesh~K.~Varanasi is with the Department of Electrical, Computer
and Energy Engineering, University of Colorado Boulder, Boulder, US,
e-mail: \protect\href{mailto:varanasi@colorado.edu}{varanasi@colorado.edu}.%
}}
\maketitle
\begin{abstract}

We study the degrees of freedom (DoF) regions of the two-user multiple-input multiple-output (MIMO) broadcast channel with a general message set (BC-CM) \textemdash that includes private and common messages \textemdash under fast fading. 
Nine different channel state knowledge assumptions \textemdash collectively known as hybrid CSIT models \textemdash are considered wherein the transmitter has either perfect/instantaneous (P), delayed (D) or no (N) channel state information (CSI) from each of the two receivers. General antenna configurations are addressed wherein the three terminals have arbitrary numbers of antennas. The DoF regions are established for the five hybrid CSIT models in which either both channels are unknown at the transmitter or each of the two channels is known perfectly or with delay. In the four remaining cases in which exactly one of the two channels is unknown at the transmitter, the DoF regions under the restriction of linear encoding strategies \textemdash also known as the linear DoF (LDoF) regions \textemdash are established.  As the key to the converse proofs of the LDoF region of the MIMO BC-CM under such hybrid CSIT assumptions, we show that, when only considering linear encoding strategies, the channel state information from the receiver with more antennas does not help if there is no channel state information available from the receiver with fewer antennas. This result is conjectured to be true even without the restriction on the encoding strategies to be linear. If true, the LDoF regions obtained for the four hybrid CSIT cases herein will also be the DoF regions for those cases. 

Many of the results of this work when specialized to even the two-message problems are new. These include the LDoF regions of the MIMO BC-CM (when one of the two channels is not known) when specialized to the MIMO BC with private messages. They also include the DoF/LDoF regions for all the hybrid CSIT models obtained by specializing the corresponding regions for the MIMO BC-CM to the case with degraded messages.

%First, we establish the DoF region of the two-user MIMO BC with only private messages. Then, by employing the approach of loosening decoding requirement of the common message, the common message can be subsumed into any one of these two private messages, and the BC-CM problem degenerates into to the BC with private messages problem. Two groups of outer bounds for the two-user BC-CM can thus be obtained from DoF region results of the two-user BC with only private messages. Interestingly, it is shown via the development of linear precoding schemes, that the outer bounds obtained in this way each of the nine hybrid CSIT assumptions are tight for the two-user BC-CM system under the matching hybrid CSIT assumption for all nine cases. Consequently, the complete DoF regions of two-user MIMO BC-CM are established under the CSIT assumptions of type `NN', `DD', `PP', `PD' and `DP'. For the `PN', `DN', `NP' and `ND' cases, the {\em linear} DoF regions are established, and conjectured to be the DoF regions.
\end{abstract}


\newpage

\section{Introduction}

Multiple-input multiple-output (MIMO) systems can provide a multiplicative gain in capacity compared to their single-input single-output (SISO) counterparts, with the multiplicative factor variously referred to as capacity pre-log, 
spatial multiplexing gain, or degrees of freedom. For example,
the point-to-point (PTP) MIMO system with $M$ transmit
antennas and $N$ receive antennas has $\min(M,N)$ degrees of freedom, i.e., its capacity grows linearly with $\min(M,N)$ in the high signal-to-noise ratio (SNR) regime \cite{Telatar1999}. Moreover, in order to achieve this rate of growth of the capacity, channel state information at the transmitter (CSIT) is not needed. 

However, CSIT plays a vital role in multi-user channels. For
example, in the two-user MIMO broadcast channel (BC), CSIT can be used to send information
along different zero-forcing beams to the two receivers simultaneously so as to not create 
interference at unintended receivers \cite{ElGamal2011}. The sum-DoF of $\min(M, N_1+N_2)$ can be achieved in this way,
where $M, N_1, N_2$ are the numbers of antennas at the transmitter and at Receivers 1 and 2, respectively;
in effect, from the DoF perspective, the availability of CSIT is the antidote that exactly neutralizes the fact that the 
receivers are distributed and non-cooperating. %It is as if the receivers fully cooperate to achieve the DoF of the resulting  point-to-point MIMO channel. 
Another example is the two-user MIMO $X$ channel, a two-transmit, two-receive 
interference network where each transmitter has two independent messages, one for each receiver, in which
CSIT can be used for zero-forcing beamforming as well as to align interference 
from the two unintended messages into the same subspace (to the extent possible) at the receiver where they are not desired \cite{jafar2008degrees,wang20162by2INGM}. The implementation of both transmitter zero-forcing
and interference alignment requires CSIT. Without CSIT, the DoF collapse 
to the extent that time-division alone is DoF-optimal \cite{huang2012degrees,Vaze2012}.

Henceforth, the term ``two-user MIMO BC" refers to the BC with general antenna configuration as defined above, and will also be referred to as the $(M,N_1,N_2)$ BC. Without loss of generality, we assume that $N_1 \geq N_2$ throughout.
%The $(M,N,N)$ BC will be called the two-user symmetric MIMO BC and the $(M,1,1)$ BC will be called the two-user MISO BC. 

Since the receivers are able to save and post-process the data, we will assume, as is commonly done in the literature, that there is perfect channel state information at the receivers (CSIR). However, the benefits of perfect and instantaneous CSIT notwithstanding, practical settings in which such CSIT can be acquired are more of an exception than the rule. 
The typical approach to obtain CSIT is to transmit pilot signals, have the receivers measure the channel state and 
send this measured channel state back to the transmitter via feedback links \cite{Jindal2006}.
In constant or slowly time-varying networks, it may be reasonable to assume that the channel
state information at the transmitter(s) acquired in this manner remains unchanged and valid 
when it is used for the subsequent transmission. 

But if the delay
between the time when the channel state information is measured and
the time when it is used at the transmitters is non-negligible compared
to the rate of channel variation, the transmitter cannot use the outdated channel state information as if it were current. A natural way to deal with this delay is to predict the current channel state using previous information and the channel time-correlation model, and then use the predicted channel state as if it were the true channel state in a scheme designed for the prefect CSIT case \cite{Caire2010}. In this scheme, the accuracy of prediction plays a significant role on the
effective (finite SNR) multiplexing gain achieved. 

When the delay is significant compared to the rate of channel variation however, even this prediction-based approach may fail in that the predicted values are poor estimates of the current channel state. In such cases, one may be better-off relying only on past channel states, even if they are independent of the current state, i.e., even if they are completely outdated. Such an approach was proposed in \cite{Maddah-Ali2012}. It was shown that even when channel fading states across symbols are independent and identically distributed (i.i.d.), in which predicting the channel state based on past information is impossible, finitely delayed channel state information is still useful in many cases (an advantage of allowing an arbitrary finite delay is that accurate estimation of the channel state becomes possible). For example, the multi-input, single-output (MISO) broadcast channel with $K$ transmit antennas and $K$ single antenna receivers can achieve a sum-DoF of $\frac{K}{1+\frac{1}{2}+...+\frac{1}{K}}$ with delayed (and accurate) CSIT, while only 1 degrees of freedom is achievable when there is no CSIT \cite{Vaze2012}. Although
$\frac{K}{1+\frac{1}{2}+...+\frac{1}{K}}$ is much smaller than $K$, which is the sum DoF of the same system under the assumption
of perfect CSIT, the scaling of sum DoF as $O(K/\log{K})$ is still significant and inspiring
compared to the no CSIT result. In \cite{Vaze2011}, the authors extend
the MISO BC results to the MIMO BC with an arbitrary number of antennas
at each terminal. An outer bound of the DoF region is provided, which
is further shown to be tight for the two-user case in \cite{Vaze2011} and for certain symmetric three-user cases (with equal numbers of antennas at all receivers) in \cite{abdoli_3user_BC_delayed_isit} by providing the respective DoF-region-optimal achievability schemes. The key idea of using delayed CSIT in  \cite{Maddah-Ali2012,Vaze2011} is that, the interference experienced by a certain receiver at a previous time is useful in the future for another receiver where that interference is a desired signal. If the transmitter re-sends a copy of that previous interference (which it can obtain using delayed CSIT feedback), not only does it benefit the other receiver where that interference is desired but it would also not cause interference at the same user again, since this user is able to cancel its influence using the saved version of the past received signal containing that interference. Thus, in this phase, transmission could be more efficient than under the no CSIT assumption. 

Besides these symmetric or homogeneous CSIT assumptions in which the availability
of CSI from all receivers are at the same level (i.e., perfect (P),
delayed (D) or no (N) CSIT), there are more general, and perhaps more commonly occurring, scenarios in which one can expect different types of CSI from different receivers due to the heterogeneity of channel variations. In \cite{Tandon2012}, the DoF region of the two-user MIMO BC is studied in which the one receiver's channel 
is known instantaneously and perfectly at the transmitter, whereas the
other receiver's channel is known to it in a delayed manner. The DoF
region in this hybrid setting, henceforth called the `PD' case\footnote{For the nine possible hybrid CSIT cases, we use a concatenation of two letters each drawn from the alphabet $\{P, D, N\}$ to denote the status of CSI from Receivers 1 and 2, in that order. For example, `PD' means that the transmitter has perfect knowledge of the first receiver's channel state and delayed knowledge of the second receiver's channel state.} is, in general, larger than that in the symmetric delayed `DD' CSIT case, and smaller than that in the symmetric perfect `PP' CSIT case. Such a phenomenon is also observed in the two-user MIMO interference channel in \cite{Kaniska-Vaze-MV:2015}. Such results on the sensitivity of even the DoF of wireless networks to the extent of availability of CSIT underscore the importance of hybrid CSIT models.
%importance of understanding hybrid CSIT models as a pre-cursor to understanding the larger question of achievable performance of wireless networks under practical settings in which there is only partial and imprecise CSIT. 

The `PN' case, in which perfect CSI is available from one receiver
and no CSI is available from the other, is more challenging. The
authors of \cite{Davoodi2014} introduce the idea of ``aligned
image sets'' and prove that the two-user MISO BC with perfect CSI from
one receiver and finite-precision CSIT from the other (hence including the `PN' case), has a maximum sum DoF of just 1.
In particular, the perfect channel knowledge
for one user at the transmitter does not help in improving the DoF beyond that of the `NN' case. The works of \cite{Mohanty-MV:KBC2013,Amuru_Tandon:SS2014,Lashgari2015} %,Rassouli2015
investigate MISO BC for more than two users under hybrid CSIT. In particular, \cite{Lashgari2015} makes significant progress that includes the exact DoF under the constraint of linear encoding strategies (known as the linear DoF, denoted LDoF) for the three-user MISO BC for all possible $3^3$ hybrid CSIT models. In spite of these advances on the MISO BC however, the generalization of the result of \cite{Davoodi2014} on the two-user MISO BC to the $(M,N_1,N_2)$ BC is, to the best of the authors' knowledge, an open problem.

Because of the special difficulty that hybrid CSIT models pose even in the two-user MIMO BC when exactly one of the two channels is not known at the transmitter, we classify the nine hybrid CSIT models as belonging to one of two types throughout this paper. Type I contains the five hybrid CSIT models $\{$`NN', `DD', `DP', `PD', `PP'$\}$\footnote{Because we assume throughout that $N_1 \geq N_2$, symmetric hybrid CSIT models, such as `PD' and `DP' or `PN' and `NP'  must be considered as two distinct models.} in which either both channels are not known or each of the two channels is known perfectly or with delay. Type II contains the other four hybrid CSIT models $\{$`ND', `DN', `NP', `PN'$\}$, in which exactly one of the two channels is not known at the transmitter. 

For Type I models, the exact DoF region of the two-user MIMO BC with private messages only (henceforth referred to as the BC-PM) have been found in the literature \cite{Vaze2011,Tandon2012,huang2012degrees,Vaze2012}. One contribution of this paper is the complete characterization of the LDoF regions of the $(M,N_1,N_2)$ BC-PM for the Type II hybrid CSIT models. A key result we obtain in this regard is a tight outer bound on the LDoF region for the `PN' (and `DN') hybrid CSIT models. 
%when $N_1 \geq N_2$.

While much work has been devoted to the study of transmitting private messages (i.e., multiple unicasting) over the broadcast channel (i.e., the BC-PM), the more general as well as the more interesting problem of simultaneous {\em groupcasting} has received much less attention. In simultaneous groupcasting, there may be exponentially many (in number of receivers) independent messages, one message desired by each distinct subset or group of receivers. 

In this paper, we study the two-user fast fading Gaussian MIMO $(M,N_1,N_2)$ BC with simultaneous two-unicasting and multicasting, i.e., the transmitter has two independent private messages intended for each of the two users, respectively, and one common multicast message which is desired at both users. %This model clearly contains all possible messages that can be transmitted over a one-transmitter, two-receiver network. 
Henceforth, we will refer to this broadcast channel with the three messages simply as the MIMO BC-CM, or as the $(M,N_1,N_2)$ BC-CM. The BC-PM is evidently a special case of the BC-CM, as is the BC with degraded messages (i.e., with a private message intended for one receiver and a common message for both receivers), denoted henceforth as the BC-DM.

The fixed two-user Gaussian MIMO BC-CM (without fading) under perfect CSIT has been extensively studied previously. An achievable scheme consisting of a linear superposition of Gaussian codewords for the common message with a dirty-paper coding (DPC) scheme for the private messages was proposed in \cite{Jindal-Goldsmith:2004}. The resulting inner bound (the DPC region) on the capacity region, was shown to be tight in certain sub-regions in \cite{Weingarten:ISIT2006}. Meanwhile, the DoF region of the two-user MIMO BC-CM, also under the perfect CSIT assumption, was obtained in \cite{Ekrem2010}. In \cite{Ekrem2010}, the generalized singular value decomposition (GSVD) was used to construct a parallel Gaussian broadcast channel so as to obtain an outer bound on the DoF region, and it was shown that that bound can be attained by an achievable scheme also based on the GSVD. As a special case of a more general result on the interference channel with general message sets, \cite{Wang2015,wang20162by2INGM} also obtain the DoF region but with a scheme based just on the singular value decomposition (SVD)\footnote{Indeed, the DoF region for the $ 2 \times 2$ network seen as two interfering BC-CMs (i.e., with two different transmitters but with common receivers) with six messages altogether is also fully established as a special case of an even more general result in \cite{Wang2015,wang20162by2INGM}.}. An outer bound based on the GVSD and relaxation of the input power in \cite{Ekrem2010} (a refinement of that in \cite{Ekrem-Ulukus2012}) is shown therein to be within an SNR-independent (but channel-dependent) constant of the DPC region of \cite{Jindal-Goldsmith:2004}, thereby providing an approximation of the capacity region within an SNR-independent additive gap. Finally, the authors of \cite{Geng-Nair-bccm-2014} prove the optimality of Gaussian inputs in Marton's inner bound to establish that the DPC region of \cite{Jindal-Goldsmith:2004} is indeed the capacity region of the two-user Gaussian MIMO BC-CM.

In what is the main result of this work, we establish the DoF regions of the two-user fast fading $(M, N_1, N_2)$ BC-CM under the Type I hybrid CSIT models and the LDoF regions for the Type II hybrid CSIT models. These results represent significant progress on the understanding of the BC-CM beyond the perfect CSIT (or `PP') setting in practically relevant scenarios, where we associate practical relevance to fast fading and the extent of availability of CSIT.
%In particular, we completely characterize the DoF region of the two-user MIMO BC-GM under the CSIT assumptions of type `NN', `DD', `PP', `PD' and `DP'. For the more challenging `PN', `DN', `NP' and `ND' cases, we establish the LDoF region, i.e., degrees of freedom assuming linear coding strategies at the transmitter. 
It is further conjectured that, for the Type II cases, the obtained LDoF regions are also the respective DoF regions.

In obtaining the outer bounds for the DoF/LDoF regions for the $(M, N_1, N_2)$ BC-CM, we demonstrate the relationship between the two-user MIMO BC-CM and the two-user BC-PM. The key idea for obtaining the outer bound on the DoF (or LDoF) region of the BC-CM is via the approach of loosening the decoding requirement of the common message so that it is decoded only at one receiver. In other words, the common message is devolved into either one or the other of the two private messages, and the outer bounds for the resulting MIMO BC-PM are then used to obtain outer bounds for the MIMO BC-CM. Remarkably, this approach works for all the nine hybrid CSIT cases, in the sense that it produces tight outer bounds for the DoF regions under hybrid CSIT cases of Type I and tight outer bounds for the LDoF regions under hybrid CSIT cases of Type II.

Then, it is shown that all the corner points of the three-dimensional DoF (respectively, LDoF) outer bound regions of MIMO BC-CM thus obtained under each of the nine hybrid CSIT assumptions have at least one zero element. The achievability proof in each case thus consists of solving one of two sub-problems: the achievability of the DoF (LDoF) region of MIMO BC-PM and the achievability of the DoF/LDoF region of MIMO BC-DM, both using linear encoding strategies. We obtain the achievability schemes for the private message MIMO BC for the Type II hybrid CSIT models (with those for Type I known in the literature) corresponding to all relevant corner points of the outer bound regions of the BC-CM. We also obtain linear achievability schemes for the MIMO BC-DM for both Type I and Type II hybrid CSIT models corresponding to all relevant corner points of the outer bound regions of the BC-CM. Any DoF-tuple in the DoF (or LDoF respectively) region of the BC-CM is then achieved using these strategies via time-sharing. Remarkably again, this high-level description of the overall strategy for obtaining the DoF/LDoF region for the MIMO BC-CM applies to each one of the nine hybrid CSIT models. In other words, in each case, it is sufficient to time-share between schemes designed for the BC-PM and the BC-DM.

Notation: $\mathbb{R}_{+}^{n}$ and $\mathbb{Z}_{+}^{n}$ denote the
set of $n$-tuples nonnegative real numbers and integers, respectively.
$(x)^{+}$ means $\max(x,0)$. ${\rm null}(A)$ denotes the nullspace
of the linear transformation $A$.

\section{System Model, DoF and LDoF}
\label{sub:BC-CM-under-hybrid}
In this section, we define the system model of the two-user MIMO BC-CM under hybrid CSIT and the DOF and LDoF metrics.

\begin{figure}[h]
\begin{centering}
\includegraphics[width=8.5cm]{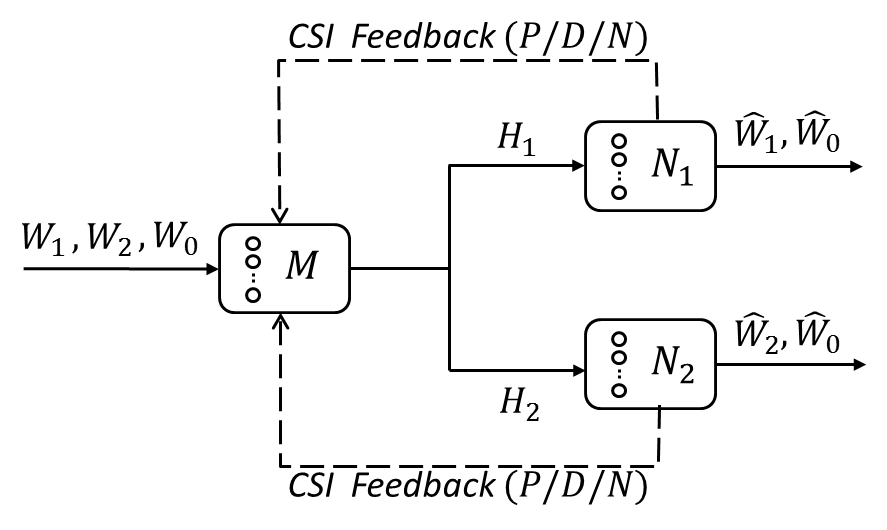}
\par\end{centering}

\caption{\label{fig:two-user-MIMO-BC-CM}2-user MIMO broadcast channel with
common message}
\end{figure}

Consider the MIMO $(M, N_1, N_2)$ Gaussian broadcast channel with arbitrary antennas
setting, i.e., the transmitter has $M$ antennas and the two users
have $N_{1}$, $N_{2}$ receive antennas, respectively. We will assume without loss of generality that $N_{1}\geq N_{2}$,
because if $N_{1}<N_{2}$, we could exchange the indexes of the two users. As is shown in Figure \ref{fig:two-user-MIMO-BC-CM}, the transmitter
has two private messages $W_{1}$ and $W_{2}$ intended for two receivers,
respectively, and one common message $W_{0}$, which is desired by
both receivers. The channel matrices $H_{1}(t) \in \mathbb{C}^{N_1 \times M}$ and $H_{2}(t) \in \mathbb{C}^{N_2 \times M}$ are i.i.d. across time and receiver indexes, and their entries are i.i.d. standard complex normal $\mathcal{CN}(0,1)$ random variables. The transmitter
can have either perfect/instantaneous (P), delayed (D) or no (N) channel
state information (CSI) available from each receiver. When considering
the delayed CSIT, without loss of generality, the delay can be taken
to be one time unit. Hence, the transmitter with delayed knowledge of 
receiver $r$'s channel knows $H_r(t-1)$ at time $t$.

The signals received at receiver $r$ ($r=1,2)$ at time $t$ is given by 
\begin{equation}
Y_{r}(t)=H_{r}(t)X(t)+Z_{r}(t),
\label{eq:model}
\end{equation}
where $X(t) \in \mathbb{C}^{M\times 1}$ is the transmitted signal at time $t$, $Z_{r}(t) \in \mathbb{C}^{N_r\times 1}$ is the additive white Gaussian noise (AWGN) vector
at receiver $r$. The channel input is subject to an average power
constraint, which is take to be $\mathrm{E}\left(X^{\dagger}(t)X(t)\right)\leq P$ for all $t$ (the superscript $\dagger$ denotes complex conjugate transpose).
For codewords occupying $t_{0}$ channel uses, we say that the rate-tuple $(R_{1},R_{2},R_{0})$ is achievable if the probabilities of
error for all three messages can be made arbitrarily small simultaneously
by choosing appropriately large $t_{0}$. The capacity region $\mathcal{C}(P)$
is then defined as the set of all achievable rate tuples $(R_{1},R_{2},R_{0})$,
while the DoF region $\mathbb{D}$ is defined as
\begin{gather*}
\begin{array}{c}
\mathbb{D}\triangleq\Biggl\{(d_{1},d_{2},d_{0})\in\mathbb{R}_{+}^{3}:\,\forall(\omega_{1},\omega_{2},\omega_{0})\in\mathbb{R}_{+}^{3}\\
\underset{x=0,1,2}{\sum}\omega_{x}d_{x}\leq\underset{\rho\rightarrow\infty}{\limsup}\left[\underset{\boldsymbol{R}(\rho)\in\boldsymbol{C}(\rho)}{\sup}\frac{\underset{x=0,1,2}{\sum}\omega_{x}R_{x}(\rho)}{\log(\rho)}\right]\Biggr\}.
\end{array}
\end{gather*}

If we restrict ourselves to linear coding strategies as defined in
\cite{Bresler2013,Lashgari2014}, in which the degrees of freedom
simply indicates the dimension of the linear subspace of transmitted
signal, we obtain the linear DoF (denoted LDoF) of the system. More
specifically, consider a linear coding scheme with block length $T$.
At time $t$, $(t=1,...,T)$, the three messages are modulated with
precoding matrix $V_{i}(t)$ $(i=1,2,0)$, respectively. The column
size of matrix $V_{i}(t)$ is equal to the number of independent information
symbols of message $W_{i}$ that will be transmitted in the entire
$T$ time slots. The signal transmitted by the transmitter at time
$t$ can be written as 
\[
S(t)=\sum_{i=0}^{2}V_{i}(t)\boldsymbol{x}_{i}^{(T)},
\]
where $\boldsymbol{x}_{i}^{(T)}\in\mathbb{C}^{m_{i}^{(T)}\times1}$
contains the entire $m_{i}^{(T)}$ information symbols. Ignoring noise,
the signal received by receiver $r$ $(r=1,2)$ is equal to $H_{r}(t)S(t)$.
Letting $V_{i}^{(T)}$ be the overall precoding matrix of message
$W_{i}$ of the entire block, and $V_i(t) \in \mathbb{C}^{M \times m_{i}^{(T)}} $ 
be its $t^{\rm{th}}$ block row (that determines the transmitted signal at time $t$, 
we have that
\[
V_{i}^{(T)}=\left[\begin{array}{c}
V_{i}(1)\\
V_{i}(2)\\
\vdots\\
V_{i}(T)
\end{array}\right].
\]
The equivalent overall channel matrix will be the block diagonal matrix
given by
\[
H_{r}^{(T)}=\left[\begin{array}{cccc}
H_{r}(1) & 0 & \cdots & 0\\
0 & H_{r}(2) & \cdots & 0\\
\vdots & \vdots & \ddots & \vdots\\
0 & 0 & \cdots & H_{r}(T)
\end{array}\right].
\]

At receiver $r$, the corresponding signal subspace is Span($H_{r}^{(T)}[V_{r}^{(T)}\ V_{0}^{(T)}]$),
the interference subspace is Span($H_{r}^{(T)}V_{r'}^{(T)}$),
where $r=1,2$ and $r'=3-r$. In order to decode the information symbols
correctly, the signal subspace and interference subspace
must be linearly independent with each other and the signal subspace must reserve
the full column rank. In other words, the following two constraints
need to be satisfied for both $r=1$ and $r=2$:
\begin{gather}
\textrm{rank}(H_{r}^{(T)}[V_{r}^{(T)}\ V_{0}^{(T)}\ V_{r'}^{(T)}])=\textrm{rank}(H_{r}^{(T)}[V_{r}^{(T)}\ V_{0}^{(T)}])+\textrm{rank}(H_{r}^{(T)}V_{r'}^{(T)})\label{eq:rank1}\\
\textrm{rank}(H_{r}^{(T)}[V_{r}^{(T)}\ V_{0}^{(T)}])=m_{r}^{(T)}+m_{0}^{(T)}\label{eq:rank2}
\end{gather}

Based on this setting, we now define the LDoF
of MIMO 2-user BC-CM.
\begin{defn}
\label{def:The-DoF-tuple}The DoF tuple $(d_{1},d_{2},d_{0})$ is
linearly achievable if there exists a sequence of linear encoding
strategies with block length of $T$, such that for each $T$ and
the choice of $m_{i}^{(T)}$$(i=1,2,0)$, $V_{i}^{(T)}$ satisfy the
decodability conditions (\ref{eq:rank1}) and (\ref{eq:rank2})
with probability 1, and
\[
d_{i}=\underset{T\rightarrow\infty}{\lim}\frac{m_{i}^{(T)}}{T}
\]
holds for all $i=1,2,0$. We also define the LDoF
region, $\mathbb{D}_{L}$, as the closure of the set of all achievable
3-tuple $(d_{1},d_{2},d_{0})$.
\end{defn}

\section{Main Results}

As stated previously, in this paper, we completely characterize
the DoF region of the 2-user MIMO BC-CM under the five
hybrid CSIT models of Type I. For the hybrid CSIT models of Type II, LDoF regions are
established. 
%These regions are conjectured to be identical to the corresponding DoF regions. 

This section is organized as follows. In Sections \ref{sub:private-PN} and \ref{sub:private-all}, we consider the two-user $(M, N_1, N_2)$ BC-PM, and establish the LDoF regions for the `PN' and `DN' in Section \ref{sub:private-PN} and the `NP' and `ND' hybrid CSIT settings in Sections \ref{sub:private-all}, respectively.  The DoF region results for the MIMO BC-PM under the other five Type I hybrid CSIT models are known in the literature. We conjecture that the four LDoF region results of Sections \ref{sub:private-PN} and \ref{sub:private-all} are also the DoF regions for the respective CSIT settings. 

In Section \ref{sub-bccm}, we establish the DoF regions under the Type I hybrid CSIT settings for the BC-CM. For the Type II cases, we generalize our results of Sections \ref{sub:private-PN} and \ref{sub:private-all} for the LDoF regions of the $(M, N_1, N_2)$ BC-PM to the $(M, N_1, N_2)$ BC-CM. These LDoF regions are also conjectured to be the DoF regions for the respective hybrid CSIT models.

\subsection{The MIMO BC-PM under hybrid CSIT of type `PN' (and `DN')}
\label{sub:private-PN}

The LDoF region for the `PN' hybrid CSIT model (which is identical to that of the `DN' model) 
for the MIMO BC with private messages is given in Theorem \ref{them:For-the-2-user}. 
Before proving that theorem, we prove the key result below.

\begin{lem}
\label{lem:For-the-2-user}For the 2-user MIMO broadcast channel with
hybrid CSIT of type `PN', if $N_{2}\leq\min(M,N_{1})$, considering
any linear coding scheme as described in Section \ref{sub:BC-CM-under-hybrid}, if $V_{1}^{(T)}$ is decodable (i.e., the symbols of message $W_1$ are all decodable) at receiver 1, we have that 
\begin{equation}
\frac{\textrm{rank}(H_{1}^{(T)}V_{1}^{(T)})}{\textrm{rank}(H_{2}^{(T)}V_{1}^{(T)})}\leq\frac{\min(M,N_{1})}{N_{2}}\label{eq:lemh1v1onh2v1}
\end{equation}
for arbitrary $T$ and $V_{1}^{(T)}$.\end{lem}
\begin{IEEEproof}
It is worth noting that we only consider the precoding matrix $V_{1}^{(T)}$
for message $W_{1}$ and its projection at both receivers in the statement of 
this lemma, so that matrix $V_{0}^{(T)}$ and $V_{2}^{(T)}$
are non-existent here in the analysis without any impact on its validity.
In other words, no symbols of message $W_{2}$
and $W_{0}$ are transmitted in the channel in the following analysis.
This setting is crucial in this proof, as we will show later.

The difficulty of the proof is that the channel matrices $H_{1}^{(T)}$
and $H_{2}^{(T)}$ are not generic matrices. They are block-diagonal matrices.
Many nice properties of generic matrices can not be directly used here.
To deal with the block-diagonal channels, we first show that there
exists a block-diagonal matrix $\hat{V}_{1}^{(T)}$, which has the
same size as $V_{1}^{(T)}$ and can be written as 
\[
\hat{V}_{1}^{(T)}=\left[\begin{array}{cccc}
\hat{V}_{1}(1) & 0 & \cdots & 0\\
0 & \hat{V}_{2}(2) & \cdots & 0\\
\vdots & \vdots & \ddots & \vdots\\
0 & 0 & \cdots & \hat{V}_{2}(T)
\end{array}\right]
\]
where $\hat{V}_{1}(i)$ , $(i=1,...,T)$ are $M\times m_{1}(i)$ matrices
with full column rank and $\sum_{i=1}^{T}m_{1}(i)=m_{1}^{(T)}$. Furthermore,
the beamformers chosen from $\textrm{Span}(\hat{V}_{1}^{(T)})$ are all
decodable at receiver 1, and the following two constraints are satisfied
\begin{gather}
\textrm{rank}(H_{1}^{(T)}V_{1}^{(T)})=\textrm{rank}(H_{1}^{(T)}\hat{V}_{1}^{(T)})=m_{1}^{(T)}\label{eq: v1plusrank1}\\
\textrm{rank}(H_{2}^{(T)}V_{1}^{(T)})\stackrel{a.s.}{\geq}\textrm{rank}(H_{2}^{(T)}\hat{V}_{1}^{(T)}).\label{eq:v1plusrank2}
\end{gather}
If such a matrix $\hat{V}_{1}^{(T)}$ exists and satisfies condition
(\ref{eq: v1plusrank1}) and (\ref{eq:v1plusrank2}), then to prove
Lemma \ref{lem:For-the-2-user}, it is sufficient to prove instead
that $\frac{\textrm{rank}(H_{1}^{(T)}\hat{V}_{1}^{(T)})}{\textrm{rank}(H_{2}^{(T)}\hat{V}_{1}^{(T)})}\leq\frac{\min(M,N_{1})}{N_{2}}$.

To begin with, we systematically construct such a $\hat{V}_{1}^{(T)}$ step-by-step
from $V_{1}^{(T)}$, and then show its aforementioned properties
hold.

Step 1: consider the first block-row in $V_{1}^{(T)}$, i.e., $V_{1}(1)$.
Let $a=\textrm{rank}(V_{1}(1))$. Then, we can express the beamformer
subspace $\textrm{Span}(V_{1}^{(T)})$ using another set of basis
vectors, such that they are column vectors of the following block
triangular matrix
\begin{equation}
V_{1,step1}^{(T)}=\left[\begin{array}{cc}
V_{1,a}(1) & 0\\
V_{1,a}(2) & V_{1,b}(2)\\
\vdots & \vdots\\
V_{1,a}(T) & V_{1,b}(T)
\end{array}\right],\label{eq:step1v}
\end{equation}
where the size of sub-matrix $V_{1,a}(i)$ is $M\times a$, and the
size of sub-matrix $V_{1,b}(i)$ is $M\times(m_{1}^{(T)}-a)$. The
way to obtain this new set of basis vectors is as follows. First,
pick any $a$ column vectors of $V_{1}^{(T)}$ whose sub-vectors
corresponding to the first time slot form a basis of $\textrm{Span}(V_{1}(1))$,
and place them as the first $a$ columns of $V_{1,step1}^{(T)}$.
This basis matrix corresponding to the first time slot is defined
as $V_{1,a}(1)$ in (\ref{eq:step1v}), and we define the sub-matrix
which contains the obtained first $a$ columns of $V_{1,step1}^{(T)}$
as $V_{1,step1}^{1:a}$. Next, for each of the rest of $(m_{1}^{(T)}-a)$
column vectors in $V_{1}^{(T)}$, which we denote generically (one by one) 
as $\left[\begin{array}{c}
\vec{v}_{1}\\
\vec{v}_{R}
\end{array}\right]$, where $\vec{v}_{1}$ contains the first $M$ rows (i.e., it corresponds
to the first time slot), and $\vec{v}_{R}$ is the remaining part. If $\vec{v}_{1}=0$,
we add this column vector directly as the next column vector of $V_{1,step1}^{(T)}$.
If $\vec{v}_{1}\neq0$, then it can be rewritten as a linear combination
of the column vectors of $V_{1,a}(1)$, say $\vec{v}_{1}=V_{1,a}(1)\vec{x}$,
where $\vec{x}$ is the $a\times1$ vector of coefficients. Then, we
add $\left[\begin{array}{c}
\vec{v}_{1}\\
\vec{v}_{R}
\end{array}\right]-V_{1,step1}^{1:a}\vec{x}$ as the next column vector of $V_{1,step1}^{(T)}$, such that its
top $M$ elements are zeros. After processing all the rest of the 
$m_{1}^{(T)}-a$ unselected vectors in $V_{1}^{(T)}$ in this way, we finally obtain the
$MT\times m_{1}^{(T)}$ dimensional new basis matrix $V_{1,step1}^{(T)}$.
The column vectors of $V_{1,step1}^{(T)}$ are guaranteed to be mutually linearly
independent since each of them contains a different
independent basis vector from $V_{1}^{(T)}$. In linearly transforming $V_{1}^{(T)}$ to $V_{1,step1}^{(T)}$, it can be shown that the subspace spanned by the beamformers in $V_{1}^{(T)}$ remains unchanged, i.e., $\textrm{Span}(V_{1}^{(T)})=\textrm{Span}(V_{1,step1}^{(T)})$. Consequently, we have that $\textrm{rank}(H_{1}^{(T)}V_{1}^{(T)})=\textrm{rank}(H_{1}^{(T)}V_{1,step1}^{(T)})$
and $\textrm{rank}(H_{2}^{(T)}V_{1}^{(T)})=\textrm{rank}(H_{2}^{(T)}V_{1,step1}^{(T)})$.

Step 2: Let $n_{1}$ be the dimension of the intersection of the beamformer
space spanned by the column vectors of $V_{1,a}(1)$ and the nullspace
of channel $H_{1}(1)$, i.e., $n_{1}=\textrm{rank}\left(\textrm{Span}\left(V_{1,a}(1)\right)\cap\textrm{null}\left(H_{1}(1)\right)\right)$.
Hence, if we only use the received signal at the first time
slot, we have that only $a-n_{1}$ independent symbols of $W_{1}$ are decodable
at receiver 1. Again, we perform a linear transformation of $V_{1,step1}^{(T)}$
to $V_{1,step2}^{(T)}$ given below
\[
V_{1,step2}^{(T)}=\left[\begin{array}{ccc}
V_{1,c}(1) & V_{1,d}(1) & 0\\
V_{1,c}(2) & V_{1,d}(2) & V_{1,b}(2)\\
\vdots & \vdots & \vdots\\
V_{1,c}(T) & V_{1,d}(T) & V_{1,b}(T)
\end{array}\right],
\]
 such that the size of $V_{1,c}(i)$ is $M\times(a-n_{1})$ and the
size of $V_{1,d}(i)$ is $M\times n_{1}$ and $\textrm{Span}(V_{1,d}(1))\subset\textrm{null}(H_{1}(1))$.
In other words, we linearly transform $V_{1,step1}^{(T)}$ such that the last $n_{1}$ column of its first block-column
will be zero-forced at Receiver 1 in the first time slot. The transformation
procedure is similar to that in Step 1, and so we omit the details for brevity. So far, the spanned beamformer subspace is still
unchanged, i.e., $\textrm{Span}(V_{1}^{(T)})=\textrm{Span}(V_{1,step2}^{(T)})$.
As a result, we still have equalities that, $\textrm{rank}(H_{1}^{(T)}V_{1}^{(T)})=\textrm{rank}(H_{1}^{(T)}V_{1,step2}^{(T)})$
and $\textrm{rank}(H_{2}^{(T)}V_{1}^{(T)})=\textrm{rank}(H_{2}^{(T)}V_{1,step2}^{(T)})$.

Step 3: We set $V_{1,d}(1)$ and $V_{1,c}(i)$, $i=2,...,T$, in $V_{1,step2}^{(T)}$
to all-zero and obtain $V_{1,step3}^{(T)}$ , i.e., 
\[
V_{1,step3}^{(T)}=\left[\begin{array}{ccc}
V_{1,c}(1) & 0 & 0\\
0 & V_{1,d}(2) & V_{1,b}(2)\\
\vdots & \vdots & \vdots\\
0 & V_{1,d}(T) & V_{1,b}(T)
\end{array}\right].
\]
The rationale is as follows: because the equivalent channel matrix $H_{1}^{(T)}$ is block-diagonal,
the value of $V_{1,d}(1)$ in $V_{1,step2}^{(T)}$ can only affect
Receiver 1 at the first time slot. Since $H_{1}(1)V_{1,d}(1)=0$,
the overall received signal at Receiver 1 is unchanged after we replace $V_{1,d}(1)$ with the all-zeros matrix 
(denoted simply as $0$). Consequently, all the symbols of $W_1$ which were decodable continue to be decodable after this replacement. Recall that the messages, i.e., $W_{0}$ and $W_{2}$ are empty, and only $W_{1}$ is transmitted over the channel.
Since $V_{1,c}(1)$ is a full column rank matrix and has no intersection
with the nullspace of $H_{1}(1)$, and $V_{1,c}(1)$ is decodable even if we just use the received signal from the first time slot\footnote{This may be not true if there are other messages in the systems, since
they will impact what Receiver 1 receives at each time slot. $V_{1,c}(1)$
may be aligned with other messages and thus it may be not decodable.}. 
Thus, no matter what value $H_{1}(i)V_{1,c}(i)$ $(i=2,...,T)$ may be, 
they can be eliminated after decoding $V_{1,c}(1)$. Consequently,
we can, without loss of generality, set $V_{1,c}(i)$ $(i=2,...,T)$ to zero instead, and the resulting
$V_{1,step3}^{(T)}$ is still decodable. As a result, we have that $\textrm{rank}(H_{1}^{(T)}V_{1,step3}^{(T)})=\textrm{rank}(V_{1,step3}^{(T)})=m_{1}^{(T)}=\textrm{rank}(H_{1}^{(T)}V_{1}^{(T)})$.

It can be shown from Lemma \ref{lem:Consider-the-matrix} and Remark \ref{rem:Also,-Lemma-} in the Appendix \ref{appendix} that $\textrm{rank}(H_{2}^{(T)}V_{1,step2}^{(T)})\stackrel{a.s.}{\geq}\textrm{rank}(H_{2}^{(T)}V_{1,step3}^{(T)})$.
%by identifying the matrices $A$  by choosing $V_{1,c}(1)=A$ and etc.

Step 4: we repeat Steps 1-3 for the rest of the block rows (successively from the second 
block row to the last one), to
finally obtain a block diagonal precoding matrix and name it $\hat{V}_{1}^{(T)}$.

From the construction of $\hat{V}_{1}^{(T)}$, we have that each of
its column vectors are decodable at receiver 1. Thus, we have that
$\textrm{rank}(\hat{V}_{1}^{(T)})=m_{1}^{(T)}$ and condition (\ref{eq: v1plusrank1})
is satisfied. We define the column rank of the $i$-th diagonal block
in $\hat{V}_{1}^{(T)}$ as $m_{1}(i)$, and we have that $\sum_{i=1}^{T}m_{1}(i)=m_{1}^{(T)}$. 

The condition (\ref{eq:v1plusrank2}) follows from the transitivity
of the inequality relation, since with each transformation of the beamforming matrix
${V}_{1}^{(T)}$ in the sequence of transformations leading to $\hat{V}_{1}^{(T)}$,
$\textrm{rank}(H_{2}^{(T)}V_{1}^{(T)})$ evolves in a monotonic non-increasing fashion to
$ \textrm{rank}(H_{2}^{(T)}\hat{V}_{1}^{(T)})$.

So far, we have the block-diagonal matrix $\hat{V}_{1}^{(T)}$ , which
is decodable at receiver 1. From (\ref{eq: v1plusrank1}) and (\ref{eq:v1plusrank2})
we have that
\begin{equation}
\frac{\textrm{rank}(H_{1}^{(T)}V_{1}^{(T)})}{\textrm{rank}(H_{2}^{(T)}V_{1}^{(T)})}\stackrel{a.s.}{\leq}\frac{\textrm{rank}(H_{1}^{(T)}\hat{V}_{1}^{(T)})}{\textrm{rank}(H_{2}^{(T)}\hat{V}_{1}^{(T)})}.\label{eq:asleqh1v1onh2v1}
\end{equation}
In order to prove (\ref{eq:lemh1v1onh2v1}), it suffices to prove
that 
\begin{equation}
\frac{\textrm{rank}(H_{1}^{(T)}\hat{V}_{1}^{(T)})}{\textrm{rank}(H_{2}^{(T)}\hat{V}_{1}^{(T)})}\leq\frac{\min(M,N_{1})}{N_{2}}.\label{eq:lemh1v1onh2v2-key}
\end{equation}

Since $H_{1}^{(T)}$, $H_{2}^{(T)}$ and $\hat{V}_{1}^{(T)}$ are
all block diagonal, the image subspaces at each receiver corresponding
to different time slot are orthogonal with each other. Thus, we have
that Span$(H_{r}^{t}\hat{V}_{1}^{(T)})$ $(t=1,...,T)$ are linearly
independent with each other for $ r=1,2$, where $H_{r}^{t}$
is the $t$-th block-row of matrix $H_{r}^{(T)}$. Since only the
values of $\hat{V}_{1}^{(T)}$ during the $t$-th time slot contribute
to Span$(H_{r}^{t}\hat{V}_{1}^{(T)})$, we have that $\textrm{rank}(H_{r}^{t}\hat{V}_{1}^{(T)})=\textrm{rank}(H_{r}(t)\cdot\hat{V}_{1}(t))$.

Because the transmitter has no CSI from receiver 2 and the channel
matrix $H_{2}(t)$ is generic, the least amount of alignment will
occur at Receiver 2. If $m_{1}(t)<N_{2}$, i.e., the number of $W_{1}$
symbols transmitted at time slot $t$ is fewer that the total available
dimension at Receiver 2, we have that $\textrm{rank}(H_{1}(t)\hat{V}_{1}(t))\overset{\textrm{a.s.}}{\leq}\textrm{rank}(H_{2}(t)\hat{V}_{1}(t))$.
In general we have that $\textrm{rank}(H_{1}(t)\hat{V}_{1}(t))\leq\textrm{rank}(\hat{V}_{1}(t))\leq m_{1}(t)$,
for all $t$. However, since message $W_{1}(t)$ needs to be decodable
at Receiver 1, we cannot have strict inequality for any $t$, for
if we did, summing over all $t$ we would have $\sum_{t=1}^{T}\textrm{rank(}H_{1}(t)\hat{V}_{1}(t))=\textrm{rank}(H_{1}^{(T)}\hat{V}_{1}^{(T)})<\sum_{t=1}^{T}m_{1}(t)=m_{1}^{(T)}$,
contradicting (\ref{eq: v1plusrank1}). Hence, we have that $\textrm{rank}(H_{1}(t)\hat{V}_{1}(t))=\textrm{rank}(\hat{V}_{1}(t))=m_{1}(t), \; \forall t $. Also, we have that $\textrm{rank}(H_{2}(t)\hat{V}_{1}(t))\leq\textrm{rank}(\hat{V}_{1}(t))=m_{1}(t)$.
Consequently, we have that $\textrm{rank}(H_{2}(t)\hat{V}_{1}(t))\overset{a.s.}{=}\textrm{rank}(H_{1}(t)\hat{V}_{1}(t))=m_{1}(t)$.
The ratio $\frac{\textrm{rank}(H_{1}(t)\hat{V}_{1}(t))}{\textrm{rank}(H_{2}(t)\hat{V}_{1}(t))}\overset{a.s.}{=}1\leq\frac{\min(M,N_{1})}{N_{2}}.$

Next, consider the case that $m_{1}(t)\geq N_{2}$. Since the number
of $W_{1}$ symbols is greater than the total available dimension
at Receiver 2 at time slot $t$, $H_{2}(t)\hat{V}_{1}(t)$ will almost
surely span the entire receiver subspace, i.e., $\textrm{rank}(H_{2}(t)\hat{V}_{1}(t))\overset{\textrm{a.s.}}{=}N_{2}$.
Meanwhile, the decodability of message $W_{1}$ requires that $m_{1}(t)\leq\min(M,N_{1})$.
Hence, we have
\begin{equation}
\frac{\textrm{rank}(H_{1}(t)\hat{V}_{1}(t))}{\textrm{rank}(H_{2}(t)\hat{V}_{1}(t))}\overset{a.s.}{=}\frac{m_{1}(t)}{N_{2}}\leq\frac{\min(M,N_{1})}{N_{2}}.\label{eq:hvhvhv}
\end{equation}
Thus, for both cases, we have that inequality (\ref{eq:hvhvhv}) is always true. 

From (\ref{eq:hvhvhv}), we obtain that 
\begin{gather*}
\textrm{rank}(H_{2}(t)\hat{V}_{1}(t))\geq\frac{N_{2}}{\min(M,N_{1})}\cdot\textrm{rank}(H_{1}(t)\hat{V}_{1}(t))\text{.}
\end{gather*}
Consequently, we have that
\begin{gather*}
\sum_{t=1}^{T}\textrm{rank}(H_{2}(t)\hat{V}_{1}(t))\geq\frac{N_{2}}{\min(M,N_{1})}\cdot\sum_{t=1}^{T}\textrm{rank}(H_{1}(t)\hat{V}_{1}(t)),
\end{gather*}
which leads to
\begin{gather*}
\sum_{t=1}^{T}\textrm{rank}(H_{2}^{t}\hat{V}_{1}^{(T)})\geq\frac{N_{2}}{\min(M,N_{1})}\cdot\sum_{t=1}^{T}\textrm{rank}(H_{1}^{t}\hat{V}_{1}^{(T)})
\end{gather*}
and
\begin{gather}
\textrm{rank}(H_{2}^{(T)}\hat{V}_{1}^{(T)})\geq\frac{N_{2}}{\min(M,N_{1})}\cdot\textrm{rank}(H_{1}^{(T)}\hat{V}_{1}^{(T)}), \label{eq:rankhvhvhv}
\end{gather}
which is the same as (\ref{eq:lemh1v1onh2v2-key}). Hence the proof is
complete.\end{IEEEproof}
\begin{rem}
In Lemma \ref{lem:For-the-2-user}, $\hat{V}_{1}^{(T)}$ is constructed
only to assist the proof of inequality (\ref{eq:lemh1v1onh2v1}).
It does not mean that by directly replacing $V_{1}^{(T)}$ in the
original system with $\hat{V}_{1}^{(T)}$, the original system can
still work. Because, $\hat{V}_{1}^{(T)}$ may conflict with $V_{2}^{(T)}$
and $V_{0}^{(T)}$, and make some messages undecodable. However, in
the analysis of Lemma \ref{lem:For-the-2-user}, this does not matter,
because the other messages are non-existent.
\end{rem}

\begin{rem}
In Lemma \ref{lem:For-the-2-user}, if $N_{2}>\min(M,N_{1})$, the
LHS of (\ref{eq:lemh1v1onh2v1}) will be almost surely equal to 1.
This follows directly from the fact that $m_{1}(t)\leq\min(M,N_{1})$
and is always less than $N_{2}$, such that $\textrm{rank}(H_{2}(t)\hat{V}_{1}(t))\overset{a.s.}{=}\textrm{rank}(H_{1}(t)\hat{V}_{1}(t))=m_{1}(t)$
is always true.
\end{rem}
Now, we are ready to give the converse proof of the LDoF region.

\begin{thm}
\label{them:For-the-2-user}For the 2-user MIMO BC-PM, if no channel state information is available
from the receiver which has fewer antennas, the availability of channel
state information (delayed or instantaneous), or lack thereof, from the other receiver
will not impact the degrees of freedom region of the system when only
considering linear coding strategies. In other words, if $N_{1}\geq N_{2}$,
the LDoF regions of the system are the same under the CSIT assumption
of type `PN', `DN' and `NN', and is given by
\begin{gather}
\frac{d_{1}}{\min(M,N_{1})}+\frac{d_{2}}{\min(M,N_{2})}\leq1.\label{eq:NN}
\end{gather}
\end{thm}
\begin{IEEEproof}
This region can be achieved by random beamforming and the simple time-division
scheme even with no CSIT. Thus, we only need to prove that (\ref{eq:NN})
is an outer bound on the LDoF region of the MIMO BC-PM if no CSI
is available from Receiver 2, which has fewer antennas.

In the case that $M\leq N_{2}$, inequality (\ref{eq:NN}) becomes
to $d_{1}+d_{2}\leq M$, which is a trivial outer bound. Thus, we
only need to consider the case that $M>N_{2}$.

Consider any linear coding strategy as described as in Section \ref{sub:BC-CM-under-hybrid}.
In this problem, since the common message $W_{0}$ is not relevant,
we remove it from all conditions. Since the total dimension of receiver
space at Receiver 2 is equal to $T\cdot N_{2}$ in the entire transmission
block of length $T$, we have that
\begin{equation}
\textrm{rank}(H_{2}^{(T)}[V_{1}^{(T)}\ V_{2}^{(T)}])\leq T\cdot N_{2}\label{eq:rankhv0}
\end{equation}
From constraints (\ref{eq:rank1}) and (\ref{eq:rank2}), we have
that
\begin{gather}
\textrm{rank}(H_{2}^{(T)}V_{1}^{(T)})+\textrm{rank}(H_{2}^{(T)}V_{2}^{(T)})\leq T\cdot N_{2}\label{eq:rankhv1}\\
\textrm{rank}(H_{1}^{(T)}V_{1}^{(T)})=m_{1}^{(T)}\label{eq:rankhv2}\\
\textrm{rank}(H_{2}^{(T)}V_{2}^{(T)})=m_{2}^{(T)}.\label{eq:rankhv3}
\end{gather}
According to Lemma \ref{lem:For-the-2-user}, we have that
\begin{equation}
\frac{N_{2}}{\min(M,N_{1})}\textrm{rank}(H_{1}^{(T)}V_{1}^{(T)})\leq\textrm{rank}(H_{2}^{(T)}V_{1}^{(T)}).\label{eq: rank1121}
\end{equation}
Together with (\ref{eq:rankhv1}), we have that
\[
\frac{N_{2}}{\min(M,N_{1})}\textrm{rank}(H_{1}^{(T)}V_{1}^{(T)})+\textrm{rank}(H_{2}^{(T)}V_{2}^{(T)})\leq T\cdot N_{2}
\]
which can be rewritten as
\begin{gather*}
\frac{\textrm{rank}(H_{1}^{(T)}V_{1}^{(T)})}{\min(M,N_{1})}+\frac{\textrm{rank}(H_{2}^{(T)}V_{2}^{(T)})}{N_{2}}\leq T.
\end{gather*}
From Definition \ref{def:The-DoF-tuple}, we have that
\begin{flalign*}
\frac{d_{1}}{\min(M,N_{1})}+\frac{d_{2}}{N_{2}} & =\underset{T\rightarrow\infty}{\lim}\frac{1}{T}\left(\frac{m_{1}^{(T)}}{\min(M,N_{1})}+\frac{m_{2}^{(T)}}{N_{2}}\right)\\
 & =\underset{T\rightarrow\infty}{\lim}\frac{1}{T}\left(\frac{\textrm{rank}(H_{1}^{(T)}V_{1}^{(T)})}{\min(M,N_{1})}+\frac{\textrm{rank}(H_{2}^{(T)}V_{2}^{(T)})}{N_{2}}\right)\\
 & \leq1.
\end{flalign*}
Thus, inequality (\ref{eq:NN}) is an outer bound on the LDoF region
of the two-user MIMO BC-PM if no CSI is available from Receiver 2, which
has fewer antennas.\end{IEEEproof}

\begin{rem}
The condition that $N_{1}\geq N_{2}$ is important in Theorem \ref{them:For-the-2-user}.
The perfect CSI from receiver 1 can in general help in reducing interference received
by Receiver 1. However, since Receiver 1 has more antennas than Receiver
2 does, it can handle more information than Receiver 2, which in turn must
be able to recover all messages if $d_{2}\neq0$. Consequently, linear
techniques such as zero-forcing message $W_{2}$ at Receiver 1 are
not necessary, and hence the CSI from Receiver 1 is not useful when considering
the LDoF region result.\end{rem}

\subsection{DoF/LDoF regions of the MIMO BC-PM under hybrid CSIT models}
\label{sub:private-all}
In this section, we again consider the two-user MIMO BC-PM. 
Of the nine hybrid CSIT settings, the DoF of five of those settings 
are known from the literature, 
two others were established by Theorem \ref{them:For-the-2-user} , and the remaining two (the `NP' and `ND' cases) are established 
in the next theorem, which also summarizes the DoF regions of all nine settings. 
These results form the basis for solving the same problems for the BC-CM.

\begin{table*}[tp]
\caption{\label{tab:DoF-region-of-BC}DoF region of the $(M,N_1,N_2)$ BC-PM under different
CSIT assumptions}

\begin{centering}
\begin{tabular}{>{\centering}p{5cm}|>{\centering}p{5cm}|>{\centering}p{5cm}}
\hline 
Perfect CSIT

(PP) & Hybrid CSIT

(PD) & Hybrid CSIT

(PN){*}\tabularnewline
\hline 
$\begin{cases}
\begin{array}{c}
d_{1}\leq N_{1}\\
d_{2}\leq N_{2}\\
d_{1}+d_{2}\leq M
\end{array}\end{cases}$ & $\begin{cases}
\begin{array}{c}
\frac{d_{1}}{\min(M,N_{1})}\leq1\\
\frac{d_{1}}{\min(M,N_{1}+N_{2})}+\frac{d_{2}}{\min(M,N_{2})}\leq1
\end{array}\end{cases}$ & $\frac{d_{1}}{\min(M,N_{1})}+\frac{d_{2}}{\min(M,N_{2})}\leq1$\tabularnewline
\hline 
\end{tabular}
\par\end{centering}

\begin{centering}
\begin{tabular}{>{\centering}p{5cm}|>{\centering}p{5cm}|>{\centering}p{5cm}}
\hline 
Hybrid CSIT

(DP) & Delayed CSIT

(DD) & Hybrid CSIT

(DN){*}\tabularnewline
\hline 
$\begin{cases}
\begin{array}{c}
\frac{d_{1}}{\min(M,N_{1})}+\frac{d_{2}}{\min(M,N_{1}+N_{2})}\leq1\\
\frac{d_{2}}{\min(M,N_{2})}\leq1
\end{array}\end{cases}$ & $\begin{cases}
\begin{array}{c}
\frac{d_{1}}{\min(M,N_{1}+N_{2})}+\frac{d_{2}}{\min(M,N_{2})}\leq1\\
\frac{d_{1}}{\min(M,N_{1})}+\frac{d_{2}}{\min(M,N_{1}+N_{2})}\leq1
\end{array}\end{cases}$ & $\frac{d_{1}}{\min(M,N_{1})}+\frac{d_{2}}{\min(M,N_{2})}\leq1$\tabularnewline
\hline 
\end{tabular}
\par\end{centering}

\begin{centering}
\begin{tabular}{>{\centering}p{5cm}|>{\centering}p{5cm}|>{\centering}p{5cm}}
\hline 
Hybrid CSIT

(NP){*} & Hybrid CSIT

(ND){*} & No CSIT

(NN)\tabularnewline
\hline 
$\begin{cases}
\begin{array}{c}
\frac{d_{1}+d_{2}}{\min(M,N_{1})}\leq1\\
\frac{d_{2}}{\min(M,N_{2})}\leq1
\end{array}\end{cases}$ & $\begin{cases}
\begin{array}{c}
\frac{d_{1}+d_{2}}{\min(M,N_{1})}\leq1\\
\frac{d_{1}}{\min(M,N_{1}+N_{2})}+\frac{d_{2}}{\min(M,N_{2})}\leq1
\end{array}\end{cases}$ & $\frac{d_{1}}{\min(M,N_{1})}+\frac{d_{2}}{\min(M,N_{2})}\leq1$\tabularnewline
\hline 
\end{tabular}
\par\end{centering}

\begin{centering}
$ $
\par\end{centering}

\centering{}{*} means the region is LDoF region.
\end{table*}

\begin{thm}
\label{thm:BC}
Let $N_{1}\geq N_{2}$. The DoF regions of the two-user
MIMO BC-PM under  the CSIT assumptions of Type I are given
in Table \ref{tab:DoF-region-of-BC}, and the LDoF regions are provided in the same table
for Type II hybrid CSIT models. We name the region of case `$X_{1}X_{2}$' where $X_{1},X_{2}\in\{P,D,N\}$,
as $\mathrm{\mathbb{D}}_{BC}^{X_{1}X_{2}}$. The label
$(X_{1}X_{2})$ denote the cases for which the corresponding region is the DoF region, whereas
$(X_{1}X_{2})^*$ is used to denote the LDoF cases in Table \ref{tab:DoF-region-of-BC}.
\end{thm}
\begin{IEEEproof}
The DoF regions for cases `PP', `DD' and `NN' are known and available
in the literature in \cite{ElGamal2011,Vaze2011,huang2012degrees,Vaze2012}, 
respectively. The DoF regions for cases `PD' and `DP' are also known from
\cite{Tandon2012}. The LDoF regions for case `PN' and `DN' were established
in Theorem \ref{them:For-the-2-user}. Next, we consider
the two remaining cases, `NP' and `ND'.

Consider the `NP' case. First, $\frac{d_{2}}{\min(M,N_{2})}\leq1$
is a trivial outer bound. Then, by adding $N_{1}-N_{2}$ extra antennas
to Receiver 2, we have a new system with $N_{1}$ antennas at both
receivers. Since adding extra antennas does not shrink the LDoF region, 
the LDoF region of this new system is an outer bound on that of the original system.
From the result for the `PN' case, we have that an outer bound on the LDoF
region of the new system is $\frac{d_{1}+d_{2}}{\min(M,N_{1})}\leq1$.
Thus, it is also an outer bound for the original system under the `NP'
assumption. Next, consider achievability. We only consider
the case that $M>N_{2}$, since otherwise, the achievable scheme is trivial since
random beamforming suffices. Since the transmitter has perfect channel
state information from Receiver 2, it is possible that it sends some
symbols of message $W_{1}$ in the null space of $H_{2}$, such that
this part can be zero-forced at Receiver 2. The maximum number of
such streams that can be zero-forced is $M-N_{2}$. To achieve any
integer-valued DoF pair $(d_{1},d_{2})$ within the
outer bound, we use the following precoding scheme. For the entire
$d_{1}$ streams of message $W_{1}$, $d_{1}^{Z}=\min(d_{1},M-N_{2})$
of them are transmitted using zero-forcing and thus will not be
received at Receiver 2. The rest of the $d_{1}-d_{1}^{Z}$ streams will be
transmitted using random beamforming. For message $W_{2}$, we transmit
all its symbols using random beamforming. Now, consider the signal
received by Receiver 1. It consists of $d_{1}+d_{2}$ independent
messages. Since $d_{1}+d_{2}\leq N_{1}$, Receiver 1 will be able
to decode all these symbols. Receiver 2 would receive $d_{1}-d_{1}^{Z}+d_{2}$
independent symbols. If $d_{1}\leq M-N_{2}$, then $d_{1}-d_{1}^{Z}+d_{2}=d_{2}\leq N_{2}$.
If $d_{1}>M-N_{2}$, then
\begin{eqnarray*}
d_{1}-d_{1}^{Z}+d_{2} & = & d_{1}-(M-N_{2})+d_{2}\\
 & = & (d_{1}+d_{2}-M)+N_{2}\\
 & \leq & N_{2}.
\end{eqnarray*}
In summary, the number of independent symbols received by Receiver
2 is also no greater than its antenna numbers. Thus, Receiver 2 will
be able to recover all these symbols. As a result, the DoF tuple $(d_{1},d_{2})$
is achieved. Since all the corner points of the outer bound are integer-valued
and thus achievable, the entire region is achievable using time sharing.

Finally, consider the `ND' case. The two outer bounds come from the
fact that the LDoF region of `ND' case is a subset of that of the LDoF region of the
`NP' case and also a subset of that of the DoF region of the `DD' case. The achievability
of `ND' case is somewhat more involved and is given later in Section
\ref{sub:`ND'-case-ofBC}.
\end{IEEEproof}

\subsection{MIMO BC-CM under hybrid CSIT}
\label{sub-bccm}
Now, let us consider the MIMO BC-CM. %The introduction of the new common message allows for multicasting simultaneously with unicasting, allowing more possible applications. 
We show in this section that obtaining the (tight) outer bounds for three-dimensional DoF region of the MIMO BC-CM is related to the two-dimensional DoF regions of the MIMO BC-PM problem under all of the nine CSIT assumptions.

\begin{table*}[tp]
\caption{\label{tab:DoF-region-of-BCCM}DoF region of two-user BCCM under different hybrid
CSIT assumptions}

\begin{centering}
\begin{tabular}{>{\centering}p{5cm}|>{\centering}p{5cm}|>{\centering}p{5cm}}
\hline 
Perfect CSIT

(PP) & Hybrid CSIT

(PD) & Hybrid CSIT

(PN){*}\tabularnewline
\hline 
$\begin{cases}
\begin{array}{c}
d_{1}+d_{0}\leq N_{1}\\
d_{2}+d_{0}\leq N_{2}\\
d_{1}+d_{2}+d_{0}\leq M
\end{array}\end{cases}$ & $\begin{cases}
\begin{array}{c}
\frac{d_{1}+d_{0}}{\min(M,N_{1})}\leq1\\
\frac{d_{1}}{\min(M,N_{1}+N_{2})}+\frac{d_{2}+d_{0}}{\min(M,N_{2})}\leq1
\end{array}\end{cases}$ & $\frac{d_{1}}{\min(M,N_{1})}+\frac{d_{2}+d_{0}}{\min(M,N_{2})}\leq1$\tabularnewline
\hline 
\end{tabular}
\par\end{centering}

\begin{centering}
\begin{tabular}{>{\centering}p{5cm}|>{\centering}p{5cm}|>{\centering}p{5cm}}
\hline 
Hybrid CSIT

(DP) & Delayed CSIT

(DD) & Hybrid CSIT

(DN){*}\tabularnewline
\hline 
$\begin{cases}
\begin{array}{c}
\frac{d_{1}+d_{0}}{\min(M,N_{1})}+\frac{d_{2}}{\min(M,N_{1}+N_{2})}\leq1\\
\frac{d_{2}+d_{0}}{\min(M,N_{2})}\leq1
\end{array}\end{cases}$ & $\begin{cases}
\begin{array}{c}
\frac{d_{1}}{\min(M,N_{1}+N_{2})}+\frac{d_{2}+d_{0}}{\min(M,N_{2})}\leq1\\
\frac{d_{1}+d_{0}}{\min(M,N_{1})}+\frac{d_{2}}{\min(M,N_{1}+N_{2})}\leq1
\end{array}\end{cases}$ & $\frac{d_{1}}{\min(M,N_{1})}+\frac{d_{2}+d_{0}}{\min(M,N_{2})}\leq1$\tabularnewline
\hline 
\end{tabular}
\par\end{centering}

\begin{centering}
\begin{tabular}{>{\centering}p{5cm}|>{\centering}p{5cm}|>{\centering}p{5cm}}
\hline 
Hybrid CSIT

(NP){*} & Hybrid CSIT

(ND){*} & No CSIT

(NN)\tabularnewline
\hline 
$\begin{cases}
\begin{array}{c}
\frac{d_{1}+d_{2}+d_{0}}{\min(M,N_{1})}\leq1\\
\frac{d_{2}+d_{0}}{\min(M,N_{2})}\leq1
\end{array}\end{cases}$ & $\begin{cases}
\begin{array}{c}
\frac{d_{1}+d_{2}+d_{0}}{\min(M,N_{1})}\leq1\\
\frac{d_{1}}{\min(M,N_{1}+N_{2})}+\frac{d_{2}+d_{0}}{\min(M,N_{2})}\leq1
\end{array}\end{cases}$ & $\frac{d_{1}}{\min(M,N_{1})}+\frac{d_{2}+d_{0}}{\min(M,N_{2})}\leq1$\tabularnewline
\hline 
\end{tabular}
\par\end{centering}

\begin{centering}
$ $
\par\end{centering}

\centering{}{*} means the region is LDoF region.
\end{table*}

\begin{thm}
\label{thm:BC-CM}Let $N_{1}\geq N_{2}$. The DoF regions of the two-user BC-CM
under the hybrid CSIT
assumptions of Type I and 
the LDoF regions for the hybrid CSIT assumptions of Type II are given in Table \ref{tab:DoF-region-of-BCCM}.
We name the region of case `$X_{1}X_{2}$' as $\mathrm{\mathbb{D}}_{BC-CM}^{X_{1}X_{2}}$,
where $X_{1},X_{2}\in\{P,D,N\}$. As in Table \ref{tab:DoF-region-of-BC}, the label
$(X_{1}X_{2})$ denote the cases for which the corresponding region is the DoF region 
whereas $(X_{1}X_{2})^*$ is used to denote the LDoF cases in Table \ref{tab:DoF-region-of-BCCM}.
\end{thm}
\begin{IEEEproof}
We give the converse proof for case `DD'. The proofs for all the other
cases follow in the same manner.

First, let us loosen the decoding requirement of the common message
$W_{0}$ and only require the first user to be able to decode it,
such that $W_{0}$ degenerates into $W_{1}$. Since loosening decoding
requirement won't hurt, the DoF region of this new system is an outer
bound of that of the original system. The new system is a MIMO
BC-PM, whose DoF region is
given in Theorem \ref{thm:BC}. Thus, we obtain the following two
outer bounds for the BC-CM system under delayed CSIT as
\begin{gather}
\frac{(d_{1}+d_{0})}{\min(M,N_{1}+N_{2})}+\frac{d_{2}}{\min(M,N_{2})}\leq1\label{eq:1}\\
\frac{(d_{1}+d_{0})}{\min(M,N_{1})}+\frac{d_{2}}{\min(M,N_{1}+N_{2})}\leq1\label{eq:2}
\end{gather}
Similarly, we can also require only the second user to be able to
decode the common message $W_{0}$ and obtain another two outer bounds
as
\begin{gather}
\frac{d_{1}}{\min(M,N_{1}+N_{2})}+\frac{(d_{2}+d_{0})}{\min(M,N_{2})}\leq1\label{eq:3}\\
\frac{d_{1}}{\min(M,N_{1})}+\frac{(d_{2}+d_{0})}{\min(M,N_{1}+N_{2})}\leq1.\label{eq:4}
\end{gather}
Combining outer bounds (\ref{eq:1}), (\ref{eq:2}), (\ref{eq:3}) and
(\ref{eq:4}) together, it can be verified that the constraints (\ref{eq:1})
and (\ref{eq:4}) are redundant. After deleting these two redundant
constraints, we obtain our final outer bound (\ref{eq:2}) and (\ref{eq:3}),
which is the same with the region $\mathrm{\mathbb{D}}_{BC-CM}^{DD}$
shown in Table \ref{tab:DoF-region-of-BCCM}.

The approach of relaxing the decoding requirement at one receiver or the other 
to get two groups of outer bounds (on DoF or LDoF, as appropriate) can be used for each of the nine different hybrid CSIT cases.
It is left to the reader to verify that the DoF/LDoF region outer bounds thus obtained are exactly as in Table \ref{tab:DoF-region-of-BCCM}.

Remarkably, the outer bounds obtained via this approach are {\em tight in every case} for the MIMO BC-CM under the corresponding hybrid CSIT assumption. 
The achievability proofs are provided later in Section \ref{sub:`PN',-`DN'-andBCCM}
to \ref{sub:`ND'-case-ofBCCM}, for which achievability schemes for the MIMO BC-DM are required. 

\end{IEEEproof}

\begin{conjecture}
The LDoF regions given in Theorem \ref{thm:BC-CM} for the hybrid CSIT models of Type II are also the DoF regions for the respective settings.
%However, extending the result to other MIMO cases does not seem to be straightforward.
\end{conjecture}

Note that the proof of this conjecture reduces to demonstrating that the LDoF region given in Theorem \ref{them:For-the-2-user} for the MIMO BC-PM is also the DoF region for the `PN' (and hence `DN') setting since all the outer bound arguments of Theorems \ref{thm:BC} and \ref{thm:BC-CM} are then valid with statements about LDoF regions replaced by the corresponding ones for DoF regions, and moreover, all the achievability schemes used to prove Theorems \ref{thm:BC} and \ref{thm:BC-CM} are linear as well.
%However, extending the result to other MIMO cases does not seem to be straightforward.

The above conjecture is true for the MISO BC-CM (when $N_1=N_2=1$), since the corresponding result was recently established for the MISO BC-PM in \cite{Davoodi2014} under the `PN' setting.

In the next section, we consider the BC-DM, which is an essential precursor to the (remaining) achievability proofs of Theorems \ref{thm:BC} and \ref{thm:BC-CM}.

\section{\label{sec:Broadcast-Channel-with-W1W0}Broadcast Channel with the Degraded Message Set ($W_{1}$, $W_{0}$)}

\begin{figure}[h]
\begin{centering}
\includegraphics[width=7.5cm]{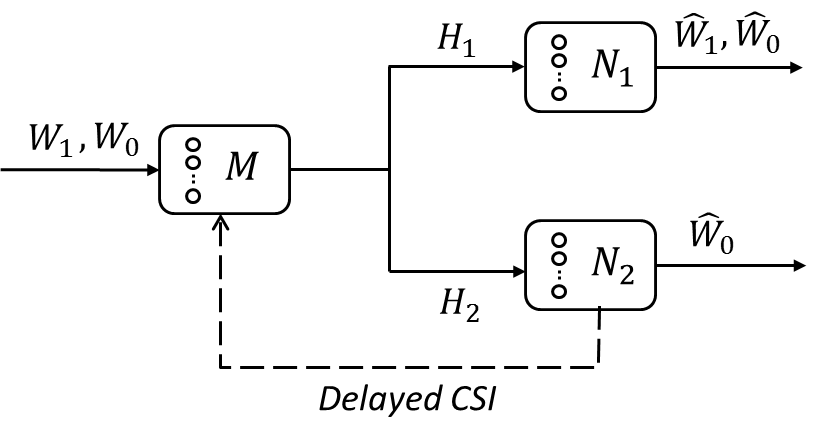}
\par\end{centering}

\caption{\label{fig:W1W0}Broadcast channel with degraded message set under
hybrid CSIT of type `DN'}
\end{figure}

Before proving the achievability of outer bounds in Theorem \ref{thm:BC}
and \ref{thm:BC-CM}, let us consider the MIMO BC-DM as shown in Figure \ref{fig:W1W0}. 
We have the same physical
structure as BC-CM (Figure \ref{fig:two-user-MIMO-BC-CM}) but now with
just two messages, $W_{1}$ and $W_{0}$. The first user
requires both messages, and the second user needs to only decode the common message
$W_{0}$. If receiver 2 has more antennas than receiver 1 does, receiver
2 will be able to recover all the messages that receiver 1 can recover.
In this case, random beamforming is optimal no matter what types of
CSIT is available, and the DoF region would simply be $d_{1}+d_{0}\leq\min(M,N_{1})$.
Thus, the $N_{1}<N_{2}$ case is trivial. 

Let us consider the case of $N_{1}\geq N_{2}$. The CSIT assumption
of type `ND' is of particular interest, since it is the key to solving
the problem under many other hybrid CSIT assumptions, as will be shown
later.
\begin{thm}
\textup{\label{thm:W1W0}If $N_{1}\geq N_{2}$, in the case that the
transmitter has no CSI from Receiver 1 but has delayed CSI from Receiver
2, i.e., hybrid CSIT of type `ND', the DoF region of the 2-user MIMO
BC-DM shown in Figure \ref{fig:W1W0}, is given by}
\begin{align}
\mathbb{D}_{BC-DM}^{ND} & =\Bigl\{(d_{1},d_{0})\Bigl|d_{1},d_{0}\geq0,\nonumber \\
 & \frac{d_{1}}{\min(M,N_{1}+N_{2})}+\frac{d_{0}}{\min(M,N_{2})}\leq1\label{eq:w1w0-1}\\
 & \frac{d_{1}+d_{0}}{\min(M,N_{1})}\leq1\Bigr\}.\label{eq:w1w0-2}
\end{align}
\end{thm}
\begin{IEEEproof}
By setting the value of $d_{2}$ in the DoF region of $\mathrm{\mathbb{D}}_{BC-CM}^{DD}$
in Table \ref{tab:DoF-region-of-BCCM} to $0$, we have an outer
bound on the DoF region of the BC-DM of Figure \ref{fig:W1W0}
under the `DD' hybrid CSIT assumption given by the inequality (\ref{eq:w1w0-1}).
This is therefore also an outer bound on the LDoF region under the `ND' assumption when no CSIT is available from Receiver 1. 
The outer bound \eqref{eq:w1w0-2} is a
simple cut-set bound. Thus, we only need to prove the achievability
of $\mathbb{D}_{BC-DM}^{ND}$.

\begin{figure}[H]
\begin{centering}
\includegraphics[width=7cm]{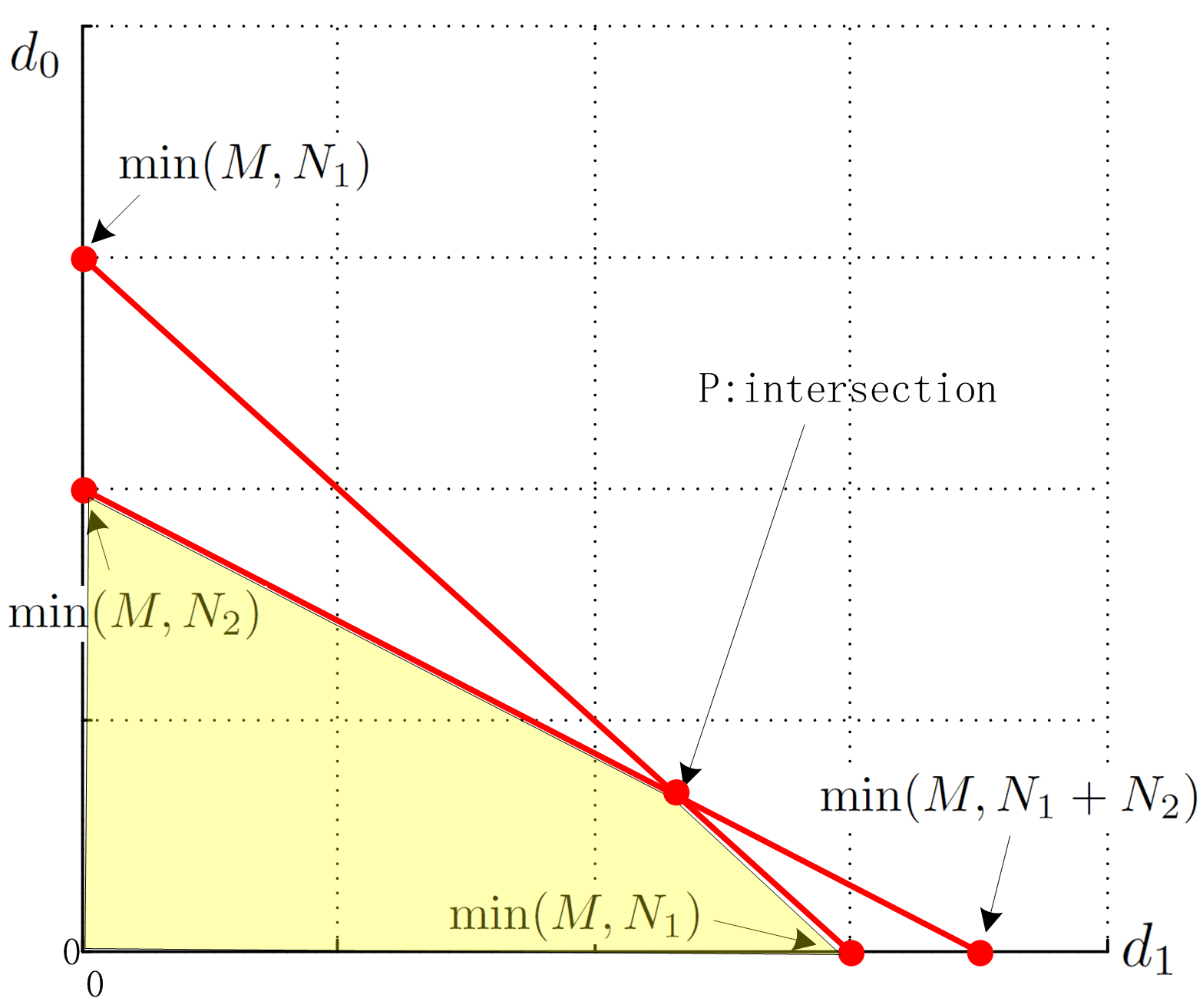}
\par\end{centering}

\caption{\label{fig:The-typical-shape}The typical shape of $\mathbb{D}_{BC-DM}^{ND}$.
There are two line boundaries (red), and the achievable region is
the quadrangle (yellow).}
\end{figure}

The typical shape of $\mathbb{D}_{BC-DM}^{ND}$ is shown in
Figure \ref{fig:The-typical-shape}. 
The two corner points $(d_{1},d_{0})=\left(\min(M,N_{1}),0\right)$
and $\left(0,\min(M,N_{2})\right)$ are trivially achievable. So to prove the
achievability of $\mathbb{D}_{BC-DM}^{ND}$, it is sufficient
to prove that point P, the intersection of the two edges, is achievable.
The entire region can then be achieved by time-sharing.

We divide the proof of achievability of corner point P into 4 cases.\\
\\
{\em Case 1}: $M\leq N_{2}$\\
In this case, the two constraints (\ref{eq:w1w0-1}) and (\ref{eq:w1w0-2})
are identical to $d_{1}+d_{0}\leq M$. This region is achievable with
random beamforming even with no CSIT, so it is trivially achieved
with CSIT of type `ND'.\\
\\
{\em Case 2}: $N_{2}<M<N_{1}$\\
In this case, the two constraints (\ref{eq:w1w0-1}) and (\ref{eq:w1w0-2})
become $\frac{d_{1}}{M}+\frac{d_{0}}{N_{2}}\leq1$ and $\frac{d_{1}+d_{0}}{M}\leq1$.
Since $N_{2}<M$, the second inequality $\frac{d_{1}+d_{0}}{M}\leq1$
is redundant. Since $(d_{1},d_{0})=(M,0)$ and $(0,N_{2})$ are achievable
with random beamforming even with no CSIT, using time-sharing, all
points in $\Bigl\{(d_{1},d_{0})\Bigl|d_{1},d_{0}\geq0,\frac{d_{1}}{M}+\frac{d_{0}}{N_{2}}\leq1\Bigr\}$
are achievable even with no CSIT. Thus, the outer bound is also trivially
achieved with CSIT of type `ND'.\\
\\
In case 1 and 2, it is easy to see that the regions are actually
equal to the corresponding DoF regions under no CSIT assumption, so
the achievability proof is trivial. For the following two cases, a
particular achievability scheme is needed to achieve corner point P. 
In this scheme, the entire transmission is divided into several
phases. The operations in specific phases are completely
different for different systems. Since the coding scheme here is almost
identical in the remaining 2 cases, we describe it with an example
for case 3, and then derive it in general in case 4.\\
\\
{\em Case 3}: $N_{1}\leq M<N_{1}+N_{2}$\\
In this case, the two constraints (\ref{eq:w1w0-1}) and (\ref{eq:w1w0-2})
become to $\frac{d_{1}}{M}+\frac{d_{0}}{N_{2}}\leq1$ and $\frac{d_{1}+d_{0}}{N_{1}}\leq1$,
and the intersection P is given by
\begin{gather*}
P=\left(\frac{M(N_{1}-N_{2})}{M-N_{2}},\frac{(M-N_{1})N_{2}}{M-N_{2}}\right).
\end{gather*}
Consider an example wherein $M=5$, $N_{1}=4$, and $N_{2}=2$, then
$P=\left(\frac{10}{3},\frac{2}{3}\right)$. To achieve this DoF pair,
we need to transmit, in 3 time slots, 10 independent symbols of private
messages $W_{1}$ to receiver one and 2 independent symbols of common
message $W_{0}$ to both receivers. Let us divide the transmission
into two phases.\\
\\
\begin{figure*}[tp]
\begin{centering}
\includegraphics[width=13cm]{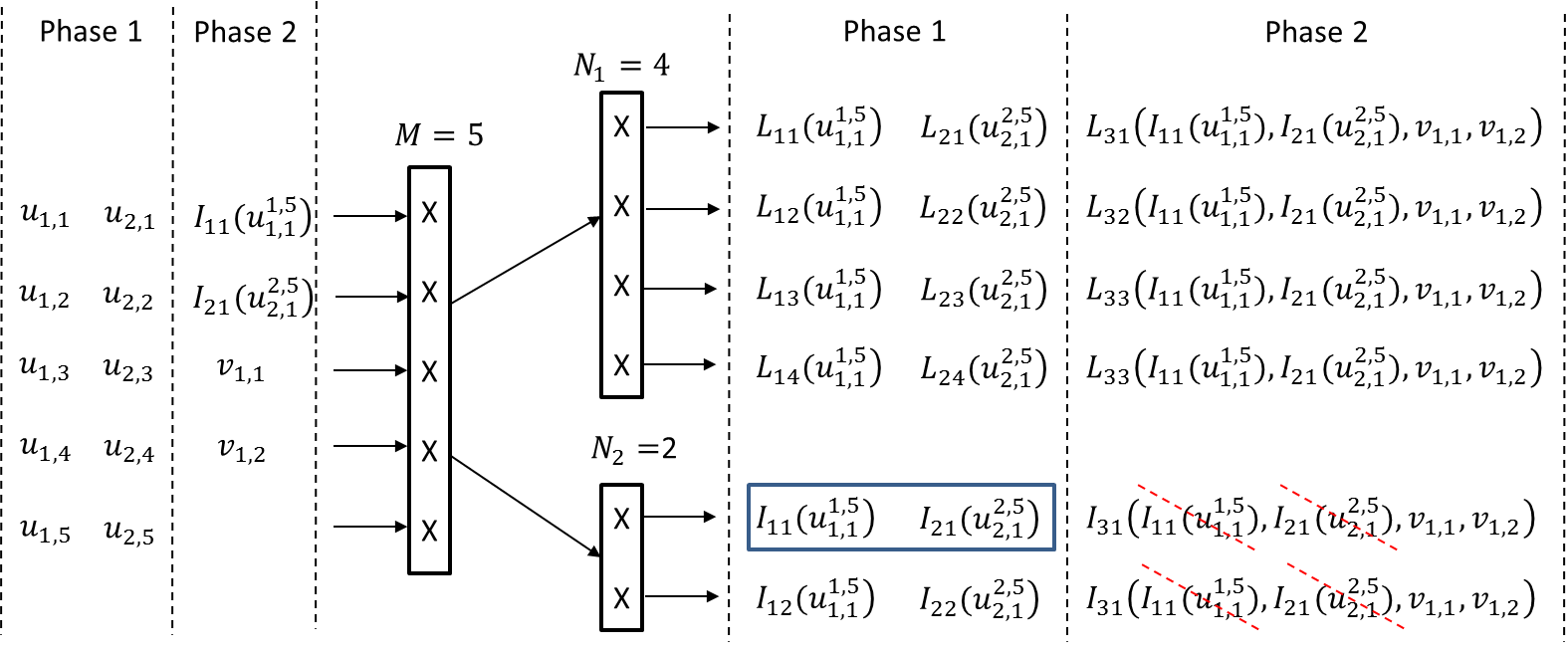}
\par\end{centering}

\caption{\label{fig:Achievable-scheme-for-case3-phase1}Achievable scheme for
case 3. In this example, phase 1 consists of 2 time slots, phase 2
consists of 1 time slot.}
\end{figure*}
\uline{Phase one} consists of $N_{1}-N_{2}=2$ time slots. At each
time slot, the transmitter sends $M=5$ independent $W_{1}$ symbols
intended for the first user through the $M=5$ transmit antennas.
Let the 5 symbols at time slot $t$ be $\{u_{t,i}\}$, where $t\in\{1,2\}$
and $i\in\{1,..,5\}$. Consider the signal received by the first user.
As is shown in Figure \ref{fig:Achievable-scheme-for-case3-phase1},
at time slot t, user one will receive $N_{1}=4$ independent linear
combinations of symbols $u_{t,1}$, $u_{t,2}$, $u_{t,3}$, $u_{t,4}$
and $u_{t,5}$, which are named $L_{t1}(u_{t,1}^{t,5})$, $L_{t2}(u_{t,1}^{t,5})$,
$L_{t3}(u_{t,1}^{t,5})$ and $L_{t4}(u_{t,1}^{t,5})$, respectively.
We have that
\begin{gather*}
\left[\begin{array}{c}
L_{t1}(u_{t,1}^{t,5})\\
L_{t2}(u_{t,1}^{t,5})\\
L_{t3}(u_{t,1}^{t,5})\\
L_{t4}(u_{t,1}^{t,5})
\end{array}\right]=H_{1}(t)\left[u_{t,1}^{*}\;u_{t,2}^{*}\;u_{t,3}^{*}\;u_{t,4}^{*}\;u_{t,5}^{*}\right]^{*}.
\end{gather*}
Similarly, user two will receive $N_{2}=2$ independent linear combinations
of symbols $u_{t,1}^{t,5}$, which are named $I_{t1}(u_{t,1}^{t,5})$
and $I_{t2}(u_{t,1}^{t,5})$. These messages are intended only for
user one, thus they are interference at user two. However, they are
still useful as explained later in phase two. We have that
\begin{gather*}
\left[\begin{array}{c}
I_{t1}(u_{t,1}^{t,5})\\
I_{t2}(u_{t,1}^{t,5})
\end{array}\right]=H_{2}(t)\left[u_{t,1}^{*}\;u_{t,2}^{*}\;u_{t,3}^{*}\;u_{t,4}^{*}\;u_{t,5}^{*}\right]^{*}.
\end{gather*}
We can observe that at each time slot, the transmitter sends 5 symbols
for user one, and user one has already obtained 4 independent equations/combinations
of them. Thus, user one only needs one more independent equation of
these 5 symbols to be able to decode them successfully.\\
\\
\uline{Phase two} consists of $M-N_{1}=1$ time slot. In this phase,
the transmitter will send $(M-N_{1})N_{2}=2$ independent symbols
$v_{1,1}$ and $v_{1,2}$ of common message $W_{0}$. Note that the
channel matrices $H_{2}(t)$ during phase one are known to the transmitter
due to the delayed CSIT assumption. As a result, the transmitter knows $I_{t1}(u_{t,1}^{t,5})$
and $I_{t2}(u_{t,1}^{t,5})$, where $t=1,2$. Since $H_{1}$ and $H_{2}$
are generic matrices and i.i.d. cross time and receiver indexes, $I_{t1}(u_{t,1}^{t,5})$
will be linearly independent with $L_{t1}(u_{t,1}^{t,5})$, $L_{t2}(u_{t,1}^{t,5})$,
$L_{t3}(u_{t,1}^{t,5})$ and $L_{t4}(u_{t,1}^{t,5})$ almost surely,
because the number of combinations is no greater than the number of
independent symbols. If the transmitter could send $I_{t1}(u_{t,1}^{t,5})$
to user one, then user one will be able be decode all 5 symbols, i.e.,
$u_{t,1}^{t,5}$.

As is shown in Figure \ref{fig:Achievable-scheme-for-case3-phase1},
in the third time slot, the transmitter will send 4 symbols, i.e.,
$I_{11}(u_{1,1}^{1,5})$, $I_{21}(u_{2,1}^{2,5})$, $v_{1,1}$ and
$v_{1,2}$, through four of its antennas. Since user one has four
receive antennas, it will be able to decode all of messages  $I_{11}(u_{1,1}^{1,5})$,
$I_{21}(u_{2,1}^{2,5})$, $v_{1,1}$ and $v_{1,2}$. Then, using $I_{t1}(u_{t,1}^{t,5})$
as well as $L_{t1}(u_{t,1}^{t,5})$, $L_{t2}(u_{t,1}^{t,5})$, $L_{t3}(u_{t,1}^{t,5})$
and $L_{t4}(u_{t,1}^{t,5})$, user one can decode message $u_{t,1}^{t,5}$ (for $t=1,2$).
In others words, user one can decode both the 2 symbols of common
message $W_{0}$ and the 10 symbols of private messages $W_{1}$.

Next, consider user two. In the third time slot, it will receive two
independent linear combinations of messages $I_{11}(u_{1,1}^{1,5})$,
$I_{21}(u_{2,1}^{2,5})$, $v_{1,1}$ and $v_{1,2}$. However, since
user two has already known\footnote{In fact, what user two knows are noisy versions of $I_{11}(u_{1,1}^{1,5})$
and $I_{21}(u_{2,1}^{2,5})$. However, noise can be neglected when considering a DoF analysis.} $I_{11}(u_{1,1}^{1,5})$ and $I_{21}(u_{2,1}^{2,5})$, it can subtract
them from the signal it receives. After removing the $I_{11}(u_{1,1}^{1,5})$
and $I_{21}(u_{2,1}^{2,5})$, it is as if user
two has 2 independent linear combinations of only $v_{1,1}$ and $v_{1,2}$.
As a result, user two can decode the 2 symbols of common message $W_{0}$.

In summary, the transmitter successfully send 10 symbols of $W_{1}$ to
user one and 2 symbols of $W_{0}$ to both users in three time slots,
i.e., the DoF $\left(\frac{10}{3},\frac{2}{3}\right)$ is achievable.\\
\\
{\em Case 4}: $M\geq N_{1}+N_{2}$\\
In this case, the two constraints (\ref{eq:w1w0-1}) and (\ref{eq:w1w0-2})
become $\frac{d_{1}}{N_{1}+N_{2}}+\frac{d_{0}}{N_{2}}\leq1$ and
$\frac{d_{1}+d_{0}}{N_{1}}\leq1$, and the intersection P is given
by
\[
P=\left(\frac{N_{1}^{2}-N_{2}^{2}}{N_{1}},\frac{N_{2}^{2}}{N_{1}}\right).
\]
The achievability scheme is almost identical with that in case 3.
We derive it in general here for case 4. Note that it is sufficient
to use only $N_{1}+N_{2}$ transmit antennas to achieve the corner
point, i.e., there are redundant antennas at the transmitter. Hence,
without loss of generality, we assume $M=N_{1}+N_{2}$ in the following
analysis.\\
\\
\uline{Phase one} consists of $N_{1}-N_{2}$ time slots. At each
time slot, the transmitter sends $M=N_{1}+N_{2}$ independent $W_{1}$
symbols intended for the user one through the $M=N_{1}+N_{2}$ transmit
antennas. Let the $N_{1}+N_{2}$ symbols at time slot $t$ be $\{u_{t,i}\}$,
where $t\in\{1,...,N_{1}-N_{2}\}$ and $i\in\{1,...,N_{1}+N_{2}\}$.
In phase one, transmitter sends out altogether $(N_{1}-N_{2})\cdot(N_{1}+N_{2})=N_{1}^{2}-N_{2}^{2}$
symbols of $W_{1}$. Consider the signal received by the first user.
At time slot t, user one will receive $N_{1}$ independent linear
combinations of symbols $u_{t,1}^{t,N_{1}+N_{2}}$, which are named
$L_{tk}(u_{t,1}^{t,N_{1}+N_{2}})$, where $k\in[1:N_{1}]$. Similarly,
user two will receive $N_{2}$ independent linear combinations of
symbols $u_{t,1}^{t,N_{1}+N_{2}}$, which are named $I_{tj}(u_{t,1}^{t,N_{1}+N_{2}})$,
where $j\in\{1,...,N_{2}\}$. 

We can observe that at each time slot, the transmitter sends $N_{1}+N_{2}$
symbols of $W_{1}$ for user one, and user one obtains $N_{1}$ independent
equations/combinations of them. Thus, user one only needs $N_{2}$
more independent equation of these $N_{1}+N_{2}$ symbols so as to
be able to decode them successfully.\\
\\
\uline{Phase two} consists of $N_{2}$ time slots. Note that the
channel matrices $H_{2}(t)$ during phase one are known to the transmitter
due to the delayed CSIT. As a result, the transmitter knows $I_{tj}(u_{t,1}^{t,N_{1}+N_{2}})$,
where $t\in\{1,...,N_{1}-N_{2}\}$ and $j\in\{1,...,N_{2}\}$. Since
$H_{1}$ and $H_{2}$ are generic matrix and i.i.d. cross time and
receiver indexes, $I_{tj}(u_{t,1}^{t,N_{1}+N_{2}})$ will be linearly
independent with each other and also with $L_{tk}(u_{t,1}^{t,N_{1}+N_{2}})$
almost surely, because the number of linear combinations is no greater
than the number of independent symbols. There are in sum $(N_{1}-N_{2})\cdot N_{2}$
messages as $I_{tj}(u_{t,1}^{t,N_{1}+N_{2}})$, and we equally divide
them into $N_{2}$ groups, where each group contains $N_{1}-N_{2}$
of them.

At each time slot of phase two, transmitter sends $N_{2}$ independent
symbols of common message $W_{0}$ and one of the $N_{2}$ groups
of $I_{tj}(u_{t,1}^{t,N_{1}+N_{2}})$, so that there are $N_{2}+(N_{1}-N_{2})=N_{1}$
symbols in total. Since user one has $N_{1}$ antennas, it is able
to decode all these $W_{0}$ symbols and $I_{tj}(u_{t,1}^{t,N_{1}+N_{2}})$.
Then, together with the previously saved $L_{tk}(u_{t,1}^{t,N_{1}+N_{2}})$
symbols, user one is able to decode all $u_{t,1}^{t,N_{1}+N_{2}}$.

Next, consider user two. At each time slot, it will receive $N_{2}$
independent linear combinations of $N_{2}$ symbols of $W_{0}$ messages
and $N_{1}-N_{2}$ symbols of $I_{tj}(u_{t,1}^{t,N_{1}+N_{2}})$.
However, since user two has already known all $I_{tj}(u_{t,1}^{t,N_{1}+N_{2}})$,
it can subtract them from the signal it receives. After removing the
contributions of $I_{tj}(u_{t,1}^{t,N_{1}+N_{2}})$ on the received signal,
it is as if user two has $N_{2}$ independent linear combinations
of $N_{2}$ symbols of $W_{0}$. As a result, user two can decode
all the symbols of common message $W_{0}$.

In summary, the transmitter successfully send $N_{1}^{2}-N_{2}^{2}$ symbols
of $W_{1}$ to receiver one and $N_{2}^{2}$ symbols of $W_{0}$ to
both receivers in $(N_{1}-N_{2})+N_{2}=N_{1}$ time slots, i.e., the
DoF $\left(\frac{N_{1}^{2}-N_{2}^{2}}{N_{1}},\frac{N_{2}^{2}}{N_{1}}\right)$
is achieved.\\
\\
All the cases together prove corner point P in Figure \ref{fig:The-typical-shape}
is achievable, and hence the DoF region described in Theorem \ref{thm:W1W0}
is achievable. 
\end{IEEEproof}

\section{\label{sec:achievability-of-outer}Achievability proof of Theorem
\ref{thm:BC} and \ref{thm:BC-CM}}

In this section, we give all the remaining achievability proofs of
Theorems \ref{thm:BC} and \ref{thm:BC-CM}.

\subsection{\label{sub:`ND'-case-ofBC}`ND' case of Theorem \ref{thm:BC}}

By comparing the region $\mathbb{D}_{BC-PM}^{ND}$ in Table \ref{tab:DoF-region-of-BC}
with $\mathbb{D}_{BC-DM}^{ND}$ given in Theorem \ref{thm:W1W0},
we find that the shapes of these two regions are exactly the same
except that one of them contains $d_{2}$ and the other one contains
$d_{0}$. Since the decoding requirement of message $W_{0}$ is higher
than that of message $W_{2}$, each symbol of message $W_{0}$ can
be thought of as a symbol of message $W_{2}$ by not requiring
receiver 1 to be able to decode it. If a DoF tuple $(d_{1},d_{0}$)
is achievable in the BC-DM,
the DoF tuple $(d_{1},d_{2})$, where $d_{2}=d_{0}$ is also achievable
in the same physical channel. In other words, the DoF region $\mathbb{D}_{BC-DM}^{ND}$
is achievable for the MIMO broadcast channel with only private
messages. The achievable scheme is the same as the scheme given in
Section \ref{sec:Broadcast-Channel-with-W1W0}.

\subsection{\label{sub:`PN',-`DN'-andBCCM}`PN', `DN' and `NN' cases of Theorem
\ref{thm:BC-CM}}

The DoF regions for these three cases are achievable by the simple
random beamforming and time-division scheme.

\subsection{`PP' case of Theorem \ref{thm:BC-CM}}

To start, we propose a precoding scheme and show that it can
achieve all the integer-valued DoF tuples within the region $\mathrm{\mathbb{D}}_{BC-CM}^{PP}$.
This scheme is also a special case/simplification of the precoding
scheme for the more general 2$\times$2 interference network with general message sets 
proposed in \cite{Wang2015,wang20162by2INGM}.

In the case that $M>N_{1}$, the null space of channel $H_{1}$ is
not empty. By transmitting symbols of $W_{2}$ using beamformers picked
from the null space of $H_{1}$, i.e., ${\rm null}(H_{1})$, we can
zero-force these symbols at Receiver 1 and thus reduce the interference
message $W_{2}$ brings to Receiver 1. The maximum number of such
independent symbols is equal to $(M-N_{1})^{+}$. Similarly, if $M>N_{2}$,
we can zero-force, maximally, $(M-N_{2})^{+}$ independent symbols
of message $W_{1}$ at Receiver 2. So, the basic idea of the precoding
scheme is that to first transmit as many symbols of private message
$W_{i}$ $(i=1,2)$ as possible in the nullspace ${\rm null}(H_{j})$
$(j=3-i)$ and then send the rest symbols of $W_{1}$ and $W_{2}$
and all the symbols of message $W_{0}$ using random beamforming.
To obtain a basis of ${\rm null}(H_{i})$, we can do a singular value
decomposition (SVD) of matrix $H_{i}$ while arranging the singular
values in non-increasing order. Then, the last $(M-N_{i})^{+}$ right-singular
column vectors, which are corresponding to singular value 0, will
form a basis of ${\rm null}(H_{i})$. 

Suppose $\vec{d}=(d_{1},d_{2},d_{0})\in\mathbb{Z}_{+}^{3}$ and $\vec{d}\in\mathrm{\mathbb{D}}_{BC-CM}^{PP}$.
Define $d_{i}^{Z}=\min\left(d_{i},(M-N_{3-i})^{+}\right)$ and $d_{i}^{R}=d_{i}-d_{i}^{Z}$,
where $i=1,2$. Here $d_{i}^{Z}$ is the number of $W_{i}$ symbols
that will be zero-forced at receiver $3-i$, and $d_{i}^{R}$ is the
number of $W_{i}$ symbols that will be transmitted using random beamforming.
Construct matrix $V_{i}^{Z}$ and $V_{i}^{R}$ such that their column
vectors are the zero-forcing beamformers and random beamformers for
message $W_{i}$, respectively. Construct matrix $V_{0}$ such that
its column vectors are random beamformers for message $W_{0}$.

Now, consider the signal received at receiver 1. Dropping the time index, we have  $Y_{1}=H_{1}\cdot(V_{1}^{Z}S_{1}^{Z}+V_{1}^{R}S_{1}^{R}+V_{2}^{R}S_{2}^{R}+V_{0}S_{0}),$
where the $S$'s are the corresponding messages. According to the
above precoding scheme, $V_{1}^{Z}$ is generated from the nullspace
${\rm null}(H_{2})$, so it is independent with channel $H_{1}$.
Meanwhile, $V_{1}^{R}$, $V_{2}^{R}$ and $V_{0}$ are all generated
randomly, and thus they are also independent with channel $H_{1}$.
Since channel $H_{1}$ is a full matrix with generic elements, the
columns of $[H_{1}\cdot V_{1}^{Z}\;H_{1}\cdot V_{1}^{R}\;H_{1}\cdot V_{2}^{R}\;H_{1}\cdot V_{0}]$
will be linearly dependent only if they have to be linearly dependent. 

Since $d\in\mathrm{\mathbb{D}}_{BC-CM}^{PP}$, we have that 
\begin{eqnarray*}
 &  & d_{0}+d_{1}=d_{0}+d_{1}^{R}+d_{1}^{Z}\leq N_{1}\\
 &  & d_{0}+d_{1}+d_{2}=d_{0}+d_{1}^{R}+d_{1}^{Z}+d_{2}^{R}+d_{2}^{Z}\leq M.
\end{eqnarray*}
Next, consider the sum of $d_{0}+d_{1}^{R}+d_{1}^{Z}+d_{2}^{R}$.
In the case that $M>N_{1}$: if $d_{2}>M-N_{1}$, we have that $d_{2}^{Z}=M-N_{1}$
and 
\begin{gather*}
d_{0}+d_{1}^{R}+d_{1}^{Z}+d_{2}^{R}+(M-N_{1})\leq M,
\end{gather*}
 which leads to $d_{0}+d_{1}^{R}+d_{1}^{Z}+d_{2}^{R}\leq N_{1}$;
if $d_{2}\leq M-N_{1}$, we have $d_{2}^{Z}=d_{2}$ and $d_{2}^{R}=0$,
such that 
\begin{gather*}
d_{0}+d_{1}^{R}+d_{1}^{Z}+d_{2}^{R}=d_{0}+d_{1}^{R}+d_{1}^{Z}\leq N_{1}.
\end{gather*}
In the case that $M\leq N_{1}$, we have $d_{2}^{Z}=0$ and $d_{2}^{R}=d_{2}$,
such that 
\begin{gather*}
d_{0}+d_{1}^{R}+d_{1}^{Z}+d_{2}^{R}\leq M\leq N_{1}.
\end{gather*}
Thus, we have that $d_{0}^{R}+d_{1}^{R}+d_{1}^{Z}+d_{2}^{R}\leq\min(M,N_{1})={\rm rank}(H_{1})$
in all cases. As a result, the column vectors of $[H_{1}V_{1}^{Z}\;H_{1}V_{1}^{R}\;H_{1}V_{2}^{R}\;H_{1}V_{0}^{R}]$
will be almost surely linearly independent with each other, since
the number of vectors is no greater than the rank of $H_{1}$. Consequently,
receiver 1 can recover all symbols of message $W_{1}$ and $W_{0}$
via linear decoding.

Following the same argument, we have that all symbols of message $W_{2}$
and $W_{0}$ are also distinguishable at receiver 2 . In other words,
the degrees of freedom $(d_{1},d_{2},d_{0})$ is achieved. It is worth
noting that only zero-forcing, which needs singular value decomposition
(SVD), and random beamforming are required in the optimal precoding
scheme.

So far, we have proved that all integer-valued degrees of freedom
tuples in $\mathrm{\mathbb{D}}_{BC-CM}^{PP}$ are achievable. It is
easy to verify that, no matter what values $M$, $N_{1}$ and $N_{2}$
are, all the corner points of the 3-D region $\mathrm{\mathbb{D}}_{BC-CM}^{PP}$
are integer-valued and thus achievable. Consequently, the entire region
of $\mathrm{\mathbb{D}}_{BC-CM}^{PP}$ is achievable using time sharing,
and we have proved that $\mathrm{\mathbb{D}}_{BC-CM}^{PP}$ is the DoF region.

Note that \cite{Ekrem2010} also studies the DoF region of a 2-user
BC-CM system with perfect CSIT , but with fixed channels and arbitrary
channel matrices (not necessarily generic), and here we are dealing
with fast fading channel where $H$'s are i.i.d. across time. The
converse proof is made much simpler than in \cite{Ekrem2010} by using
the basic strategy of loosening decoding requirement. Also, in the
proof of achievability, we propose a relatively simpler scheme, which
needs singular value decomposition (SVD) instead of generalized singular
value decomposition (GSVD).

\subsection{`DD' case of Theorem \ref{thm:BC-CM}}

\begin{figure}[H]
\begin{centering}
\includegraphics[width=7cm]{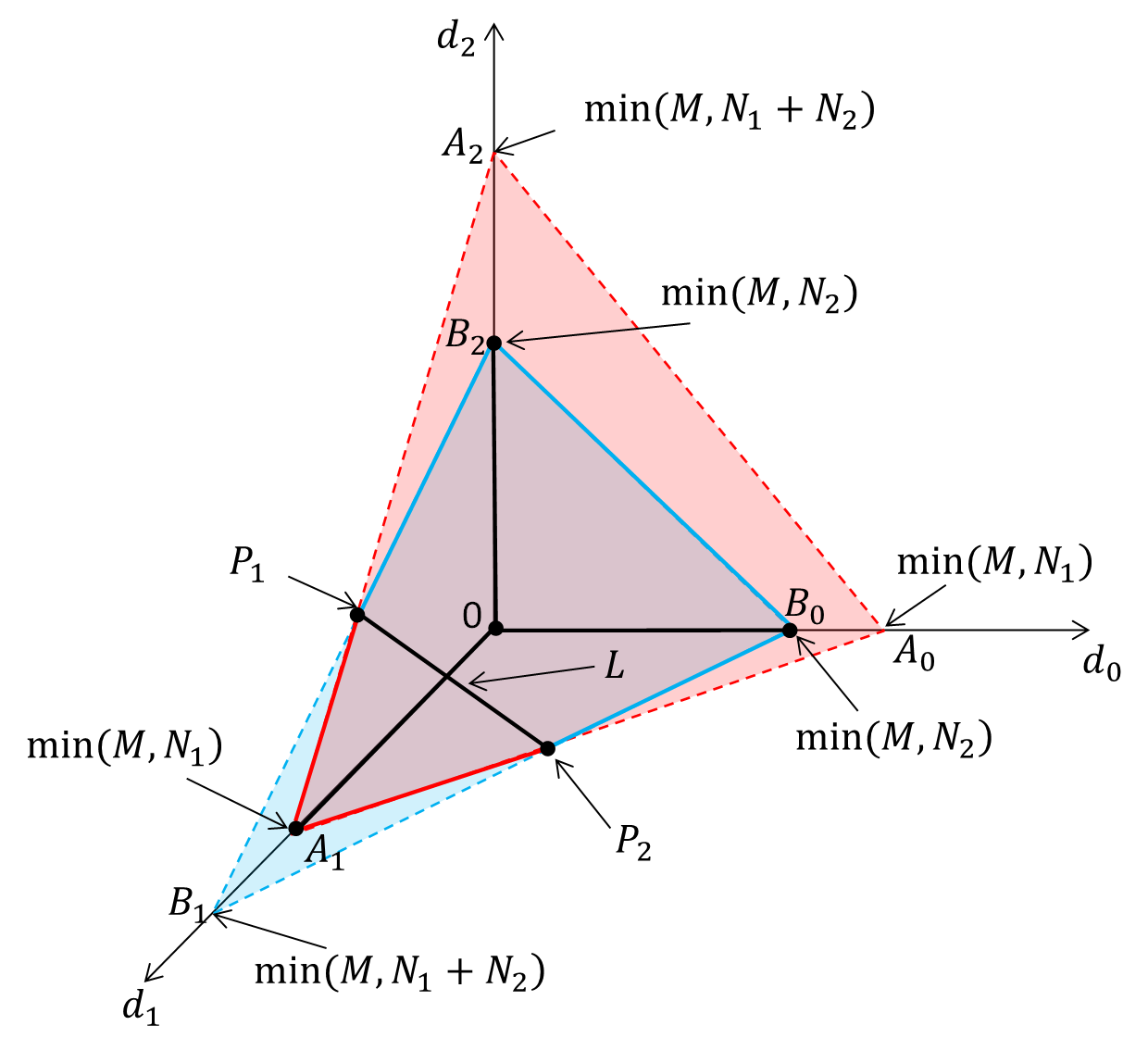}
\par\end{centering}

\caption{\label{fig:The-typical-shape-1}The typical shape of $\mathrm{\mathbb{D}}_{BC-CM}^{DD}$
. The two planes/constraints, spanned by points $(A_{1},A_{2},A_{0})$
and by $(B_{1},B_{2},B_{0})$, intersect at line $L$. The final DoF
region is a pentahedron, whose vertices are 0, $A_{1}$, $B_{0}$,
$B_{2}$, $P_{1}$ and $P_{2}$.}
\end{figure}

Observe that $\mathrm{\mathbb{D}}_{BC-CM}^{DD}$ is a three-dimensional
pentahedron. The typical shape of $\mathrm{\mathbb{D}}_{BC-CM}^{DD}$
is shown in Figure \ref{fig:The-typical-shape-1} above. There are
five non-trivial corner points on the pentahedron's boundary, and
it is sufficient to prove these corner points are achievable because
the entire region can then be achieved using time-sharing.

The three corner points $(d_{1},d_{2},d_{0})$ on the axes, i.e.,
$A_{1}=\left(\min(M,N_{1}),0,0\right)$, $B_{2}=\left(0,\min(M,N_{2}),0\right)$
and $B_{0}=\left(0,0,\min(M,N_{2})\right)$, can be achieved even
with no CSIT. Hence, they are trivially achieved with delayed CSIT.
The corner point $P_{1}$ lies in the plane $\{(d_{1},d_{2},d_{0})|d_{0}=0\}$.
It is actually the exact same corner point as that in the MIMO BC-PM
with delayed CSIT. Thus, it is achievable using the transmission scheme
proposed in \cite{Vaze2011}. The corner point $P_{2}$ lies in
the plane $\{(d_{1},d_{2},d_{0})|d_{2}=0\}$. This point is exactly the
same corner point as the one we considered in Section \ref{sec:Broadcast-Channel-with-W1W0},
so it is achievable using the transmission scheme described there. Since $P_{2}$ can be achievable
under `ND' CSIT assumption, it is also achievable under the `DD' CSIT
assumption using the same coding scheme.

Hence, all the corner points are shown to be achievable, and thus
the entire region $\mathrm{\mathbb{D}}_{BC-CM}^{DD}$ is achievable
using time-sharing.

\subsection{The `PD' case of Theorem \ref{thm:BC-CM}}

Again, in the `PD' case. There are two non-trivial corner points
which are not on the axes. One of them lies in the plane $\{(d_{1},d_{2},d_{0})|d_{0}=0\}$.
It is actually the exact same corner point as that in the MIMO BC
with `PD' CSIT. Thus, it is achievable using the transmission scheme
introduced in \cite{Tandon2012}. The other corner point lies in the
plane $\{(d_{1},d_{2},d_{0})|d_{2}=0\}$. It is actually the exact
same corner point as the one we considered in Section \ref{sec:Broadcast-Channel-with-W1W0},
and it is achievable even under `ND' CSIT assumption, so it is also
achievable under the `PD' CSIT assumption using the same coding scheme.

Hence, all the corner points are shown to be achievable, and thus
the entire region $\mathrm{\mathbb{D}}_{BC-CM}^{PD}$ is achievable
using time-sharing.

\subsection{\label{sub:`DP'-case-of}`DP' case of Theorem \ref{thm:BC-CM}}

For the case of `DP', although the shape of region seems to be symmetric
with that of case `PD', the two non-trivial corner points are still
in the plane $d_{0}=0$ and $d_{2}=0$, since $N_{1}\geq N_{2}$.
The corner point in the plane $d_{0}=0$ is again achievable using
the scheme introduced in \cite{Tandon2012}. The corner point in the
plane $d_{2}=0$ is equal to $\left(\min(M,N_{1})-\min(M,N_{2}),0,\min(M,N_{2})\right)$.
To achieve this point, we use following scheme. First, transmit the
common message $W_{0}$ using random beamforming. Hence it will occupy
$\min(M,N_{2})$ dimensions each at the two receivers. Then, since the transmitter
has perfect knowledge of channel $H_{2}$, zero-forcing some or all
of private message $W_{1}$ at Receiver 2 is possible. Because we
have $\min(M,N_{1})-\min(M,N_{2})\leq(M-N_{2})^{+}$, which is
the rank of the null-space of $H_{2}$, we can actually zero-force
all $\min(M,N_{1})-\min(M,N_{2})$ streams of private message $W_{1}$
at Receiver 2. Consequently, Receiver 2 is able to recover the $\min(M,N_{2})$
streams of common message $W_{0}$, and Receiver 1 is able to recover
the altogether $\min(M,N_{1})$ streams of message $W_{1}$ and $W_{0}$,
since the number of independent streams at neither receiver is great
than its number of antennas.

It is worth noting that it requires `DP' CSIT to achieve the corner
point in the plane $d_{0}=0$, however, `NP' CSIT is enough to achieve
the corner point in the plane $d_{2}=0$.

\subsection{`NP' case of Theorem \ref{thm:BC-CM}}

From Section \ref{sub:`DP'-case-of}, we can obtain that, under `NP'
CSIT assumption, the region $d_{1}+d_{0}\leq\min(M,N_{1})$, $d_{0}\leq\min(M,N_{2})$
is achievable. By loosening the decoding requirement of part of message
$W_{0}$ and only require receiver 2 to be able to decode them, this
part of $W_{0}$ will degenerate into message $W_{2}$. Since
loosening the decoding requirement won't hurt, we have that $d_{1}+(d_{2}+d_{0})\leq\min(M,N_{1})$,
$(d_{2}+d_{0})\leq\min(M,N_{2})$ is also achievable, which is the
same as region $\mathrm{\mathbb{D}}_{BC-CM}^{NP}$.

\subsection{\label{sub:`ND'-case-ofBCCM}`ND' case of Theorem \ref{thm:BC-CM}}

Again, the two non-trivial corner points are in the plane $d_{0}=0$
and $d_{2}=0$. The one in the plane $d_{0}=0$ is the same as the
corner point given in Section \ref{sub:`ND'-case-ofBC} and is thus achievable.
The other one in the plane $d_{2}=0$ is the same as the corner point
given in Section \ref{sec:Broadcast-Channel-with-W1W0} and is thus achievable.

\section{Conclusion}

In this paper, we study the DoF of MIMO BC
with private and common messages (BC-CM) under all possible hybrid CSIT assumptions. 
For the five Type I hybrid CSIT assumptions, we obtained the DoF regions and 
for the remaining four Type II CSIT assumptions we obtain the LDoF regions. 
The outer bounds on the DoF region for the Type I CSIT assumptions are obtained
as extensions of the respective DoF regions for the 
MIMO BC with private messages (BC-PM), which are known from previous literature.
The outer bounds on the LDoF region for the Type II CSIT assumptions are obtained
from the respective outer bounds on the LDoF region for the MIMO BC with private 
messages (BC-PM), which in turn are also obtained in this paper.

As the most important converse proof of this paper, we show in Theorem \ref{them:For-the-2-user}
that if no channel information is available from the receiver which has fewer antennas, the availability of 
channel state information from the other receiver will not impact the DoF region of the 2-user MIMO
BC-PM when only considering linear encoding strategies. In other words, channel state information 
from the receiver with more antennas does not help if no channel state information is available from the 
receiver with fewer antenna. The converse proof of the LDoF region for the MIMO BC-PM and BC-CM under Type II 
hybrid CSIT assumptions all follow from this theorem. For the achievability proof, it is shown that every corner point 
of the MIMO BC-CM DoF or LDoF regions is either a corner point of the BC-PM or a corner point of the BC with degraded
messages (BC-DM). Thus, the achievability of the BC-CM DoF/LDoF region is decomposed into series of sub-problems. 
An important such sub-problem is the MIMO BC-DM with private message to Receiver 1 (with greater number of receive antennas than Receiver 2)
and a common message under hybrid CSIT assumption in which Receiver 1's channel is unknown at the transmitter and Receiver 2's channel is known with delay.
For this setting, we propose a two-phase coding scheme to show that the outer bound on its LDoF region is tight. This sub-problem is shown to be the foundation of the achievability proof for the DoF/LDoF region of the MIMO BC-CM under multiple CSIT assumptions.

%It is shown that every corner point of the BC-CM DoF or LDoF regions is either a corner point of the BC-PM or a corner point of the BC with degraded messages (BC-DM). As an important sub-problem, we obtain an outer bound on the DoF region for the MIMO BC-DM ($W_{1}$ and $W_{0}$)  under hybrid CSIT assumption of type `ND', and then propose a two-phase coding scheme to show that the outer bound is tight.  Together with several previously known DoF results for the BC-PM under Type I CSIT assumptions and LDoF regions obtained herein for  the BC-PM under Type II CSIT assumptions, we give DoF-optimal schemes for the MIMO BC-CM under all nine different CSIT assumptions and prove the converses in the case of CSIT of type `NN', `DD', `PP', `PD' and `DP'. For the remaining four scenarios, i.e., case `PN', `DN', `NP' and `ND', we establish the linear degrees of freedom (LDoF) region, i.e., assuming linear coding strategies at the transmitters.  We conjecture that the obtained LDoF regions are the same with the DoF regions for these four cases. 

The results of this work give rise to several interesting future research directions. One such direction is to prove our conjecture that the LDoF regions obtained in this paper are indeed the DoF regions in each of the four hybrid CSIT models in the two-user MIMO BC-PM setting, as well as in the more general two-user MIMO BC-CM. In fact,  it is sufficient to prove that Theorem \ref{them:For-the-2-user} holds despite removing the restriction of linear encoding strategies, since all the other converses follow that case of MIMO BC-PM as they do in this paper but with that restriction in place.  Another direction for future research is generalizing the results of this paper for the private messages only setting to the three-user MIMO BC with a general antenna configuration. Furthermore, the DoF or even the LDoF region of the MIMO broadcast channel with a general message set, consisting of seven different messages (one for each subset of receivers where it is desired) even in the perfect CSIT is an intriguing open problem. 
%Going from the perfect CSIT setting to hybrid CSIT models for the three-user MIMO BC (with private general message sets is expected to present formidable technical challenges.

\appendices{}
\section{\label{appendix}}
\begin{lem}
\label{lem:Consider-the-matrix}Consider the matrix $X=\left[\begin{array}{cc}
A & B\\
C & D
\end{array}\right]$, where $A$, $B$, $C$ and $D$ are all sub-matrices whose sizes
satisfied the concatenation requirement. If $X$ has full column rank,
then 
\begin{equation}
\textrm{rank}\left(\left[\begin{array}{cc}
H_{1}A & H_{1}B\\
H_{2}C & H_{2}D
\end{array}\right]\right)\stackrel{a.s.}{\geq}\textrm{rank}\left(\left[\begin{array}{cc}
H_{1}A & 0\\
0 & H_{2}D
\end{array}\right]\right),\label{eq:ABCD0}
\end{equation}
where $H_{1}$ and $H_{2}$ are two generic matrices independent with
each other and also with $X$.\end{lem}
\begin{IEEEproof}
First, we prove that 
\begin{equation}
\textrm{rank}\left(\left[\begin{array}{cc}
H_{1}A & H_{1}B\\
H_{2}C & H_{2}D
\end{array}\right]\right)\stackrel{a.s.}{=}\textrm{rank}\left(\left[\begin{array}{c}
H_{1}A\\
H_{2}C
\end{array}\right]\right)+\textrm{rank}\left(\left[\begin{array}{c}
H_{1}B\\
H_{2}D
\end{array}\right]\right)\label{eq:ABCD1}
\end{equation}
Then, from the trivial facts that 
\begin{gather*}
\textrm{rank}\left(\left[\begin{array}{c}
H_{1}A\\
H_{2}C
\end{array}\right]\right)\geq\textrm{rank}\left(H_{1}A\right)\\
\textrm{rank}\left(\left[\begin{array}{c}
H_{1}B\\
H_{2}D
\end{array}\right]\right)\geq\textrm{rank}\left(H_{2}D\right).
\end{gather*}
and 
\[
\textrm{rank}\left(\left[\begin{array}{cc}
H_{1}A & 0\\
0 & H_{2}D
\end{array}\right]\right)=\textrm{rank}\left(H_{1}A\right)+\textrm{rank}\left(H_{2}D\right),
\]
we have inequality (\ref{eq:ABCD0}).

Consider (\ref{eq:ABCD1}). It indicates that the $\textrm{Span}\left(\left[\begin{array}{c}
H_{1}A\\
H_{2}C
\end{array}\right]\right)$ and $\textrm{Span}\left(\left[\begin{array}{c}
H_{1}B\\
H_{2}D
\end{array}\right]\right)$ are linearly independent with each other almost surely. Suppose there
exist a vector, $\vec{v}$, which belongs to both $\textrm{Span}\left(\left[\begin{array}{c}
H_{1}A\\
H_{2}C
\end{array}\right]\right)$ and $\textrm{Span}\left(\left[\begin{array}{c}
H_{1}B\\
H_{2}D
\end{array}\right]\right)$ . Then, there exist two non-trivial column vectors $\vec{x}$ and
$\vec{y}$, such that
\[
\vec{v}=\left[\begin{array}{c}
H_{1}A\\
H_{2}C
\end{array}\right]\vec{x}=\left[\begin{array}{c}
H_{1}B\\
H_{2}D
\end{array}\right]\vec{y}.
\]
Then, we have $H_{1}(A\vec{x}-B\vec{y})=0$ and $H_{2}(C\vec{x}-D\vec{y})=0$.
Consequently, $A\vec{x}-B\vec{y}=0$ or $\in\textrm{null}(H_{1})$,
and $C\vec{x}-D\vec{y}=0$ or $\in\textrm{null}(H_{2})$. Since $X$
has full column rank, $A\vec{x}-B\vec{y}=0$ and $C\vec{x}-D\vec{y}=0$
can not be true at the same time. If we select $\vec{x}$ and $\vec{y}$
such that $A\vec{x}-B\vec{y}=0$, we need that $C\vec{x}-D\vec{y}$
be zero-forced by $H_{2}$. However, since $H_{2}$ is a generic matrix
independent of $A$, B, C and D, the probability that $C\vec{x}-D\vec{y}$
falls in the nullspace of $H_{2}$ is almost surely zero. Similarly,
if we select $\vec{x}$ and $\vec{y}$ such that $A\vec{x}-B\vec{y}\in\textrm{null}(H_{1})$,
it is almost sure that $C\vec{x}-D\vec{y}\notin\textrm{null}(H_{2})$.
Consequently, such a vector $\vec{v}$ does not exist almost surely.
Thus, we have (\ref{eq:ABCD1}).\end{IEEEproof}
\begin{rem}
The high block-dimension extension, i.e., $X$ in the form of $N\times N$
($N>2$) sub-blocks, of Lemma \ref{lem:Consider-the-matrix} follows
in the extra same way. We omit the detailed proof due to simplicity.
\end{rem}

\begin{rem}
\label{rem:Also,-Lemma-}Also, Lemma \ref{lem:Consider-the-matrix}
can be extended straightforwardly to the following case and higher
block-dimension, under the same problem setting. 
\[
\textrm{rank}\left(\left[\begin{array}{ccc}
H_{1}A_{1} & H_{1}A_{2} & H_{1}A_{3}\\
H_{2}B_{1} & H_{2}B_{2} & H_{2}B_{3}\\
H_{3}C_{1} & H_{3}C_{2} & H_{3}C_{3}
\end{array}\right]\right)\stackrel{a.s.}{\geq}\textrm{rank}\left(\left[\begin{array}{ccc}
H_{1}A_{1} & 0 & 0\\
0 & H_{2}B_{2} & H_{2}B_{3}\\
0 & H_{3}C_{2} & H_{3}C_{3}
\end{array}\right]\right).
\]
The detailed proof is left to the reader, if interested.
\end{rem}

\bibliographystyle{unsrt}
\bibliography{BC_2user_CM_DelayedCSIT}

\end{document}